\documentclass[sn-basic]{sn-jnl}% Math and Physical Sciences Reference Style
\jyear{2022}%

\theoremstyle{thmstyleone}%
\theoremstyle{thmstyletwo}%

\theoremstyle{thmstylethree}%
\usepackage{caption}
\makeatletter
\newcommand\specialcaption[1]{%
  \@namedef{the\@captype}{#1}%
  \addtocounter{\@captype}{-1}\caption}
\makeatother

\begin{document}

\title[Self-organized criticality in neuronal network]{Self-organized criticality in a mesoscopic model of excitatory-inhibitory neuronal populations by short-term and long-term synaptic plasticity}


\author*[1]{\fnm{Masud} \sur{Ehsani}}\email{masud.ehsani@mis.mpg.de}

\author[1,2]{\fnm{J\"{u}rgen} \sur{ Jost}}\email{jjost@mis.mpg.de}

\affil*[1]{ \orgname{Max Planck Institute for Mathematics in Sciences}, \orgaddress{\street{Inselstr.22}, \city{Leipzig}, \postcode{04103}, \state{Saxony}, \country{Germany }}}

\affil[2]{\orgname{Santa Fe Institute}, \orgaddress{\street{1399 Hyde Park Rd}, \city{Santa Fe}, \postcode{NM 87501}, \country{United States }}}


\abstract{ In \cite{Ehsani22}, we have shown that the dynamics of an
    interconnected  population of
    excitatory and inhibitory spiking neurons wandering around  a  Bogdanov-Takens (BT)
    bifurcation point can generate  the
    observed scale-free avalanches at the population level and the highly variable
    spike patterns of individual neurons. These characteristics match
    experimental findings for spontaneous intrinsic activity in the brain. In this paper, we address the
    mechanisms causing the system to get and remain near this BT point. 
We propose an effective stochastic neural field model which captures the
dynamics  of the mean-field model. We show how the network tunes itself
through local long-term synaptic plasticity by STDP and short-term synaptic
depression to be close to this bifurcation point.  The mesoscopic model that we derive matches the directed percolation model at the absorbing state phase transition.}


\maketitle

\section{Introduction}

 Neural networks as complex dynamical systems with many degrees of freedom
 varying over different time scales can be seen as a self-tuning system that
 attains a dynamical regime where  the system can carry out its task. On the
 other hand, spontaneous intrinsic activity of cortical neural assemblies in
 absence of any information processing task can be perceived as a substrate
 for the neural dynamics which can give us insights into the preferred
 dynamical regime and the goal of self-organization processes. Dynamic and
 functional characteristics of spontaneous activity are connected to the
 structural architecture of the brain as well as the ongoing self-organization
 process.  Experimental findings on different temporal and spatial resolutions
 highlight the scale-free characteristic of spontaneous activity. When
 recorded by coarse-grained methods like EEG and MEG, spontaneous brain
 activity shows nested oscillations with a power spectrum that indicates
 scale-free properties, i.e. $P(f) \propto 1/f^{\beta}$(\cite{Hansen01,Miller09, Hardstone12}). Microelectrode recordings of smaller cortical
 regions show activity in the form of avalanches with power-law distributions
 of size and duration in different setups such as cultured slices of rat
 cortex (\cite{Beggs03}), awake monkeys
 (\cite{Petermann09}), in cerebral cortex and hippocampus of
 anesthetized, asleep, and awake rats ( \cite{Ribeiro10}) and visual cortex of anesthetized cat ( \cite{Hahn10}). 
 
One explanation for the scale-free characteristics is that cortical networks operate near a critical second-order phase transition (\cite{Chialvo04, Tagliazucchi13}). On one hand, close to the edge of an active-inactive phase transition, local populations of neurons in the cortex would be in an idle state ready for processing information, but at the same time away from overactivation.
On the other hand, close to the phase transition, an order parameter, a
macroscopic state different from the inactive state, comes into existence. The
emergence of a macroscopic mode of activity acts as the coordinator of
individual neurons which are enumerated in quantity and prone to various kinds
of noise to produce a cooperative large-scale activity (\cite{Chialvo10}.
Macroscopic states of activity, in turn, enslave the individual neurons. These active states in the spontaneous mode in the absence of meaningful information processing tasks can be seen as either a random sequence of active neurons or activation of a sequence of already inscribed patterns of co-activity in the connectivity map.  
 Mean-field models of neural dynamics can be used as the first step in the coarse-grained description of the neural network. Rate models can accurately describe neural networks in all-to-all or sparsely connected networks. The appropriateness of mean-field solutions for the all-to-all network in the limit of large size should be clear.
On the other hand, in a sparsely connected network, correlations among the input to two different neurons, beyond the average population rate, are assumed to be negligible, and correspondingly also the magnitude of cross-correlation between spike trains of neurons is small.
Therefore, in the population of neurons in low firing regimes and in the sparsely connected network in which neurons fire with Poisson statistics but asynchronously, the total population rate dynamics can be modeled by a mean-field equation. The asynchronous state of population rate in the EI population can itself be oscillatory.
 In \cite{Brunel99} and \cite{Brunel00}, a sparsely connected network with synaptic delay between inhibitory feedback and the excitatory rate has been studied using a Fokker-Planck formalism for the evolution of the membrane voltage in the asynchronous state. \cite{Brunel08} also studied fast global oscillations in neural networks operating at a low rate.
The above-mentioned rate equations, also called Neural Mass Models, can be
extended to the continuum limit. Continuum neural field models as nonlinear
integrodifferential equations with the integral kernel representing the
connectivity strength between different neural populations have been
introduced by \cite{Wilson72,Wilson73}. Neural field models
can show wave propagation in terms of a front solution in a bistable network
(see \cite{Amari77,Ermentrout98}), propagating pulses in an
excitable medium (see \cite{Pinto01}), and spatially localized oscillations
and spiral waves in the oscillatory regime of a local EI population
(\cite{Troy07}). They can also have localized bump solutions \cite{Pinto012}
and spatially periodic patterns called Turing patterns (\cite{Ermentrout79}). Weakly nonlinear analysis, singular perturbation methods, symmetric bifurcation theory, homogenization theory, and stochastic processes are the analytic tools for investigating these patterns of activity. (See \cite{Bressloff11}, for a comprehensive review.)

  Mean-field solutions can only exist in the limit of $N \to \infty$ and are
  based on neglecting correlations among neurons and finite-size
  effects. Stochastic neural networks have been proposed to account for
  fluctuations in the asynchronous state of firing and studying their
  correlations and finite-size effects in different studies
  (\cite{Ginzburg94,Meyer02,Soula07,Touboul15}). These networks model
  microscopic neural dynamics as a Markov process based on the assumption that
  neural dynamics is Markovian. Truncation at the first moment of the
  mentioned Markov process is compatible with the mean-field equation. By
  truncation of the higher moments based on the assumption that pairwise
  correlations are of order $1/N$ and $p$-moments are of order $1/N^{p-1}$, we
  can derive closed-form equations for higher moments of activity of neurons
  or sub-populations of neurons.  The main assumption is that in the
  asynchronous state auto-correlations are of order $1/N$. 
  \cite{Bressloff00} utilized system size expansion of the master equation to
  systematically truncate the moment hierarchy based on the system size.
  Moreover, the stochastic version of the continuum neural field has been
  discussed in \cite{Buice07} and  \cite{Bressloff09}.
  
 In general, the mesoscopic description of dissipative systems in which the
 flow and fluctuation of energy are not based on the equilibrium fluctuations,
 such as general reaction-diffusion systems, is very complicated. The absence
 of detailed balance in microscopic dynamics and the fluctuation-dissipation
 theorem would make a straightforward phenomenological mesoscopic approach
 like the Langevin dynamics for systems close to equilibrium
 impossible. However, for the non-equilibrium system which has a steady-state
 close to a critical point and shows generic scale invariance, an effective
 dynamical description in terms of a mesoscopic field equation is in general possible. In the critical state, the perturbations in the system spread in all length and time scales of the system with power-law distributions of size and duration.
   
In this case, we can write down the dynamics of the system out of equilibrium
in terms of the microscopic master equation, like for interacting particle
systems. Using the well-known coherent path integral procedure first
introduced in the context of reaction-diffusion equations by Doi and Peliti (
\cite{Doi76,Peliti85}) to form a non-Hermitian bosonic Hamiltonian and a
continuum field representation, internal fluctuations are automatically taken
into account. By using the perturbation approach and the renormalization group method, we can study the scale invariance and critical behavior of the model.  
Subsequently, the field theory representation in terms of a path integral over
stochastic paths can be translated to the Langevin equation with an appropriate noise term using Janssen-Dedominicis functional representation.
 
  Buice and Cowan used  a coherent path formalism and the Doi-Peliti functional representation  to translate the microscopic master equation to a path integral representation for activity fields (\cite{Buice07}) . One advantage of their method is that the study of scale invariance at criticality in the functional representation is possible. They proposed that the stochastic neural field equation for the excitatory system at a critical point can be written in the form of Langevin's description of directed percolation. Benayoun et. al.  proposed a stochastic model of spiking neurons which matches the Wilson-Cowan mean field in the limit of infinite system size that shows scale-free avalanches in the balanced state in which sum of excitation and inhibition is much larger than the net difference between them (\cite{Benayoun10}). Under symmetry condition on weights this makes the Jacobian to have negative eigenvalues close to zero in the balanced state. \cite{Cowan13} used the method of path integral representation in the stochastic model of spiking neurons supplemented by anti-Hebbian synaptic plasticity as the self-organizing mechanism. Their network possesses bistability close to the saddle-node bifurcation point which is the origin of the avalanche behavior in the system.

To tune the system at the critical point, many modeling approaches and adaptive mechanisms have been suggested during the decades of research on the critical brain hypothesis. A SOC model for a non-conservative neuronal model that attracted much attention is introduced by \cite{Levina09, Levina07}.
In addition to self-organization by short-term depression in synapses which is
also used in  \cite{Peng13} and \cite{Santo17}, self-organization by other control parameters like degree of connectivity or synaptic strength has been studied. Adaptive rewiring of asymmetric synaptic connections with fixed strength ($J_{ij}= \pm 1)$ by the average input correlation was introduced in the spin models of neural networks (\cite{Bornholdt03, Rybarsch14}). In these works, the authors have claimed that insertion and deletion of the links based on adaptive rewiring regulate the network toward criticality by tuning the branching ratio to unity. 

In line with the methods used in those,  \cite{Meisel09} introduced a self-organizing neural network by STDP.  In \cite{Brochini16}, self-organization in stochastic spiking neuron model by short term plasticity of the gain function instead of synaptic weights is introduced. \cite{Benayoun10} proposed a model composed of stochastic single neurons which shows avalanche dynamics in the regime of closely balanced input.

 In \cite{Ehsani22}, we showed that the Bogdanov-Takens bifurcation point of the mean-field equations for dynamics of an sparse homogenous excitatory and inhibitory population of spiking neurons with conductance-based synaptic currents is the operating point of the system producing the characteristic spontaneous activity in the form of scale-free avalanches.  
 Here, we consider the self-tuning of the system at this critical point. The self-organizing parameter in our network is the balance of opposing forces
 resulting from 
 the activities of inhibitory and excitatory populations,  and the
 self-organizing mechanisms are long-term synaptic plasticity through the
 mechanism of  Spike Timing Dependent Plasticity (STDP) and homeostatic short-term depression of the synapses.  
The former tunes the overall strength of excitatory and inhibitory pathways to be close to a balanced regime of these currents and the latter, which is based on the finite amount of resources in brain areas, acts as an adaptive mechanism that tunes micro populations of neurons subjected to fluctuating external inputs to attain the balance in a wider range of external input strengths.  
For analytical analysis of STDP on average weight connections, we use the inhomogenous Poisson neuron assumption that has been studied in \cite{kistler00} and \cite{Burkitt07}. 
Under general conditions on inhibitory and excitatory STDP kernels, i.e. negative integral of the excitatory and positive integral of the inhibitory STDP kernels, learning results in a balanced internal state. This condition on kernels leads to stabilization of rates as also discussed in \cite{kempter01}.
We use the model of \cite{Markram97} for short-term depression of the excitatory synapses which has been studied vastly (see for example \cite{kistler99}). 
	
Using the Poisson firing assumption, we propose a microscopic Markovian model
which captures the internal fluctuations in the network due to the finite size
and matches the macroscopic mean-field equation by coarse-graining. Near the
critical point, a phenomenological mesoscopic model for excitatory and
inhibitory fields of activity is possible due to the time scale separation of
slowly changing variables and fast degrees of freedom.  We will show that the
mesoscopic model corresponding to the neural field model near the local
Bogdanov-Takens bifurcation point matches the Langevin description of the directed percolation process. Tuning the system at the critical point can be achieved by coupling fast population dynamics with slow adaptive gain and synaptic weight dynamics, which make the system wander around the phase transition point. Therefore, by introducing short-term and long-term synaptic plasticity, we have proposed a self-organized critical stochastic neural field model.

\section{Materials and Methods}
\subsection{Neuron model}
We use an integrate and fire neuron model in which the change in the membrane voltage of the neuron receiving time dependent synaptic current $i(t)$  follows 

  \begin{equation} \label{eq:1}
	C \dfrac{d v(t)}{dt} = g_{Leak}( v_{Leak} - v(t)) + i(t)
	\end{equation}

for $v(t)<v_{th}$ . When the membrane voltage reaches $v_{th} = -50mv$, the  neuron spikes and immediately its membrane voltage resets to $v_{rest}$ which is equal to $v_{Leak} = -65mv$.

In the following, we want to concentrate on a model with just one type of inhibitory and one type of excitatory synapses, which can be seen as the average effect of the two types of synapses. We can write the synaptic inhibitory and excitatory current as 	
		
	\begin{equation}\label{eq:2}
	\begin{aligned}
	 i(t) = & g_{inh}(t) * ( V_{Rinh} - v(t)) + g_{exc}(t) * (V_{Rexc} - v(t))
	\end{aligned}
	\end{equation}

$V_{Rinh}$ and $V_{Rexc}$ are the reverse potentials of excitatory and inhibitory ion channels, and based on experimental studies we choose values of $-80mv$ and $0mv$ for them respectively. 
 $g_{Inh}(t)$ and  $g_{Exc}(t)$ are the conductances of inhibitory and
 excitatory ion channels. These conductances are changing by the inhibitory
 and excitatory input to the cell. Each spike of a presynaptic inhibitory or
 excitatory neuron $j$ to a postsynaptic neuron $k$ that is received by  $k$
 at time $t_0$  will change the inhibitory or excitatory ion channel 
 conductance of the postsynaptic neuron for $t>t_0$ according to :
 \begin{align}
	& g_{Inh}^k(t) = w_{kj} * g_{0}^{inh} * exp(-\dfrac{t-t_0}{\tau_{syn}^{inh}}) \nonumber \\
	&g_{Exc}^k(t) = w_{kj} * g_{0}^{exc} * exp(-\dfrac{t-t_0}{\tau_{syn}^{exc}}) \label{eq:3}
	\end{align}	
 Here we assume that the rise time of synaptic conductances is very small compared to other time scales in the model and therefore, we modeled the synaptic current by a decay term with synaptic decay time constant $\tau_{syn}$ which we assume to be the same value of $5ms$ for both inhibitory and excitatory synapses.

\subsection{Network architecture}
 
In the remainder of this work, in the simulation, we consider a population of
$N_{Exc} =2*10^4$ and $N_{Inh} = 0.25*N_{Exc}$ inhibitory spiking neurons with
conductance-based currents introduced in this section. Each excitatory neuron
in the population is randomly connected to $k_{EE} = \dfrac{N_{Exc}}{100}=200$
excitatory and $k_{EI} = \dfrac{k_{EE}}{4}$ inhibitory neurons and each
inhibitory neuron is connected to $k_{IE} = k_{EE}$ and $ k_{II} = \dfrac{k_{EE}}{4} $
excitatory and inhibitory neurons, respectively. The weights of excitatory synaptic
connections are in a range that $10-20$ synchronous excitatory spikes suffice
to depolarize the target neuron to the level of its firing threshold when it is initially at rest at the time of input arrival. Weights are being drawn from a log-normal probability density with low variance. Therefore, approximately $O(\sqrt{k_{EE}})$ spikes are adequate for firing. Assuming homogeneity in the population as we have discussed in the introduction we can build a mean-field equation for the excitatory and inhibitory population in this sparse network, assuming each neuron receives input with the same statistics.  

\subsection{Avalanche regime of activity as desired operating point of the system}
In \cite{Ehsani22}, we have investigated dynamics of excitatory and inhibitory (EI) sparsely connected populations of spiking leaky integrate neurons with conductance-based synapses.
We have seen that close to the Bogdanov-Takens bifurcation point of the mean
field equation, the output firing of the population is in the  form of avalanches with scale free size and duration distribution.   This matches the characteristics of low firing spontaneous activity in the cortex. By linearizing gain functions and excitatory and inhibitory nullclines, we  approximated the location of the BT bifurcation point. This point in the control parameter phase space corresponds to the internal balance of excitation and inhibition and a slight excess of external excitatory input to the excitatory population. Due to the tight balance of average excitation and inhibition currents, the firing of the individual cells is fluctuation-driven. Around the BT point, the spiking of neurons is a Poisson process and the population average membrane potential of neurons is approximately at the middle of the operating interval $[V_{rest}, V_{th}]$. Moreover, the EI network is close to both oscillatory and active-inactive phase transition regimes. 

At equilibrium, population rates satisfy a system of fixed point
  equations of the form :
 \begin{align} 
   & \rho_{I} =  g^I(\rho _{I} , c_{EI}\rho_{E} + c_{II}\rho_{I} + d\rho_{Ext}^I) - z_0 \nonumber \\
   & \rho_{E} =  g^E(\rho _{I} , c_{EE}\rho_{E} + c_{EI}\rho_{I} + d\rho_{Ext}^E) -z_0 \label{eq:4}
\end{align}
where $c_{xy} = c k_{xy} w_{xy}(V_{R_y} - \langle V_x \rangle)$. $k_{xy}$ is the average
number of synaptic connections between neurons of population $y$ to neurons in
population $x$ with average strength of $w_{xy}$. $z_0$ is a constant that 
   depends on $V_{rest}, V_{th}$, the maximal rates and the standard deviation of the 
   input.  $V_{R_y}$ is the reverse
potential level of a neuron of type $y$ and $ \langle V_x \rangle$ is the average potential level of neurons in population $x$ that can be written in fluctuation driven the firing regime as :
   
   \begin{equation} \label{eq:5}
<V_{x}> = \dfrac{g_{L}V_{L} + g_{exc}^0 w_{xE} \rho_E \tau V_{Rexc} + g_{inh}^0 w_{xI} \rho_I \tau V_{Rinh}  }{g_{L} + g_{exc}^0 w_{xE} \rho_E \tau+ g_{inh}^0 w_{xI} \rho_I \tau}
\end{equation}	
   
  $\tau$ is the synaptic current decay time constant, $g_{L}$ , $g_{exc}$ and $g_{inh}$ are the baseline conductances of leaky, excitatory and inhibitory ion gates, respectively. We took $w_{EE}$ and $\rho_{Ext}^E $
as control parameters and analyzed solutions to the Eq.\ref{eq:4}. By substituting nonlinear gain functions with their corresponding linearization in the Poisson firing regime, we showed that the low BT point is located close to the
matching condition for the $y$-intercept and the slopes of the linearized
nullclines, which are written as:
  \begin{align} 
  & c_{EE}^* = \dfrac{c_{EI}c_{EI}}{c_{II}} \nonumber\\
  & \rho_{Ext}^{E^*} = \dfrac{c_{EE}^*}{c_{IE}}(\rho_{Ext}^I-d) +d \label{eq:6}
 \end{align}
 where $d$ is a constant equal  to $\dfrac{g_{L}(V_{rest} - V_{th}) }{\tau* g_{exc}^0*(V_{th}-V_{R_{exc}})}$.

\section{Results}

\subsection{Long term synaptic plasticity by STDP tunes synaptic weights close to the balanced state}

A typical neuron in the cortex has $10^3-10^4$ synaptic connections with
$80\%$ of them of excitatory type and $20\% $ of inhibitory type. On the other
hand, even in the resting state, neurons on average have a non-zero firing
rate with an average rate of $1Hz$ and their spike trains are very noisy with
exponential inter-spike interval distribution indicating that the spiking of
individual neurons is  a Poisson point process. Yet another experimental fact
about synaptic strength between neurons states is that usually, $10-20$
presynaptic synchronous spikes suffice to bring a typical neuron to the firing
threshold. If we take $\tau_m = 20ms$ as the membrane potential decay time
constant, then during this time window a typical neuron receives $20-200$
excitatory spikes, which are enough for the neuron to periodically spike at a
very high rate. To avoid this, the inhibitory input in this time window should
largely cancel the excitatory current. Therefore, for the average currents to
maintain the average membrane potential below the threshold in order to avoid
a high firing state and produce high variability in the spike trains,
inhibitory and excitatory currents should be balanced. Dynamical balance of
excitation and inhibition ensures a low level of activity, i.e., an asynchronous firing state. 
In the following, we present a synaptic plasticity rule which tunes the
average synaptic weights to the balanced state. We will analyze and simulate a
network in which neurons will adapt their connections according to the
Spike-timing dependent plasticity (STDP) paradigm  (\cite{Gerstner96}), which
provides a foundation for temporal coding. We derive an equation for the evolution of the average and the variance of weights between excitatory and inhibitory neurons during the plasticity period.
In STDP, the weight of a connection is modified depending on the time interval
between pairs of pre- and post-synaptic spikes. For every pair, the weight of the synapse changes according to the  equations
\begin{align}
\Delta w (\Delta t) = \begin{cases}
f_{+}(w) K_+ (\Delta t) 
\ \ \text{if } \Delta t \geq 0
\\
-f_{-}(w) K_- (\Delta t) 
\ \ \text{if } \Delta t < 0
\end{cases} \label{eq:7}
\end{align}
where $\Delta t = t_{post} - t_{pre}$ is the time difference between the
postsynaptic spike and the presynaptic one. The functions $f_+$ and $f_-$
model the dependence of the weight change on the current value of the synaptic
weights. $K_+$ and $K_-$, called STDP kernels, usually are decaying functions
of time which reflects the fact that closer pre- and post-synaptic spikes
generate stronger weight  changes. Usually, we model the kernels by a single
exponential such as $K_+ = A_+  e^{-\frac{\lvert \Delta t \rvert} {\tau_{s_+}}}$ and $ K_-
= A_- e^{-\frac{\lvert \Delta t \rvert}{\tau_{s_-}}}$. As it is evident from equation (\ref{eq:7}), 
when the postsynaptic neuron fires after the presynaptic neuron, the strength
of the connection increases and it decreases for the opposite temporal
order. We assume the same type of the STDP rule for both inhibitory and
excitatory connections although with different kernels. In the following, we
suppose that the dependence of STDP on the synaptic weight is negligible and
therefore replace the functions $f_+$ and $f_-$ by a constant which is then
absorbed into the kernels. In this case, we have to assume a saturation level for the maximum strength of the synapses, $w_{max}^E$ and $w_{max}^I$.

STDP changes synaptic weights on a very slow time scale compared to firing
dynamics of the neurons, therefore, during a time period of $ [t, t + \Delta
t]$ where $\Delta t$ is long in comparison with the inter-spike time interval but small enough that the change in the weight $w_{ij}$ of the synapse from neuron $j$ to neuron $i$ is infinitesimal, we can write:
\begin{align}
 \Delta w_{ij}= & \int_t^{t+\Delta t } \int_0^{\infty} S_j (s) S_{i}(s+ \delta) f_+ (w_{ij}) K_+(\delta) d\delta  ds \nonumber \\   &+ \int_t^{t+\Delta t } \int_0^{\infty} S_i (s) S_{j}(s- \delta) f_- (w_{ij}) K_-(\delta) d\delta  ds  \label{eq:8} 
\end{align}
where $S_i(t)$ and $S_{j}(t)$ are the spike trains of the presynaptic and the postsynaptic neurons. Assuming that during this period, the firing rate of the output neuron is constant on average and there exist many pre- and post-synaptic spikes, we can write  the mean change in the incoming synaptic weights to the neuron $i$ as,
\begin{equation} \label{eq:9} 
 \dfrac {\langle \Delta w_{ij} \rangle_j }{\Delta T} = \int_0^{\infty} \langle S_j (s) S_{i}(s+ \delta)\rangle _j  K_+(\delta) d\delta    -   \int_0^{\infty} \langle S_j (s) S_{i}(s- \delta)\rangle _j   K_-(\delta) d\delta  \\
\end{equation}
We want to investigate the evolution of the synaptic weights in the EI
population in an asynchronous irregular state. Therefore, we assume that in
the regime of the spontaneous activity neurons are firing as a Poisson
process. Moreover, to estimate the cross-correlation of the pre- and the  post-synaptic spike train we  argue that the excitatory input to the cell has a positive correlation with preceding spikes in the target neuron. The magnitude of this excess correlation  depends on the weight of the synapse and it is restricted to the time window before the firing of the postsynaptic neurons. With this in mind, we use the following approximation introduced in \cite{cateau03} and \cite{vanRossum00} for the cross-correlations of spike trains to account for the causal contributions of presynaptic spikes to the postsynaptic ones of a synapse with the strength $w_{i}$:
\small
\begin{align} 
& \langle S_{pre}^E (s) S_{post}^E (s+ \delta)\rangle = \rho_{pre}^E \rho_{post}  +  \rho_{pre}^E w_i(V_{Rexc} - V_{i}) \gamma^E (\delta) \nonumber \\
& \langle S_{pre}^I (s) S_{post}^E (s+ \delta)\rangle = \rho_{pre}^I \rho_{post}  -  \rho_{pre}^I  w_i ( V_{i} - V_{Rinh}) \gamma^I (\delta)  \label{eq:10}
\end{align}	
\normalsize
 Here $V_i$ is the voltage level of the postsynaptic neuron. As the second terms in both equations encode the excess correlation (anticorrelation) of the  presynaptic excitatory(inhibitory) input preceding the firing at the postsynaptic neuron, we set  $\gamma ^I (\delta)=\gamma^E (\delta) = 0 $ for $\delta < 0 $. For positive values of $\delta$, this function which is independent of the rates and the weights of the synapses, encodes the causal effect of the presynaptic spike which arrives $\delta$ units of time before firing of the postsynaptic neuron. Therefore, it is a decaying function of $\delta$. Moreover, we have assumed the dependence on the weight of the synapse to be of a linear form, which is a good approximation in the regime of small synaptic weights.
 Inserting the above approximation and labeling the  STDP kernels of EE  and
 IE synapses as $K^{E}$ and the STDP kernels of II and EI synapses as $K^{I} $, we can write  the evolution of the average excitatory and inhibitory synaptic strength to the neuron $i$ as 
 \small
  \begin{align}
& \dfrac {d\langle w_{ij}^E \rangle}{dt} = \langle  \rho_{j}^E \rangle \rho_{i} (\overline{K}_+^E - \overline{K} _-^E )+ \langle \rho_{j}^E \rangle \langle w_{ij}^E \rangle (V_{Rexc} - \langle  V_{i} \rangle ) \overline{K_+^E \gamma^E} \nonumber \\
& \dfrac {d\langle w_{ik}^I \rangle}{dt} = \langle  \rho_{k}^I \rangle \rho_{i} (\overline{K}_+^I -\overline{K} _-^I ) - \langle \rho_{k}^I \rangle \langle w_{ik}^I \rangle (\langle  V_{i} \rangle - V_{Rinh} ) \overline{K_+^I \gamma^I} \label{eq:11}
\end{align}
  \normalsize
\begin{figure}
			\centering
			\includegraphics[width=1\linewidth,trim={0cm, 0cm, 0cm, 0cm},clip]{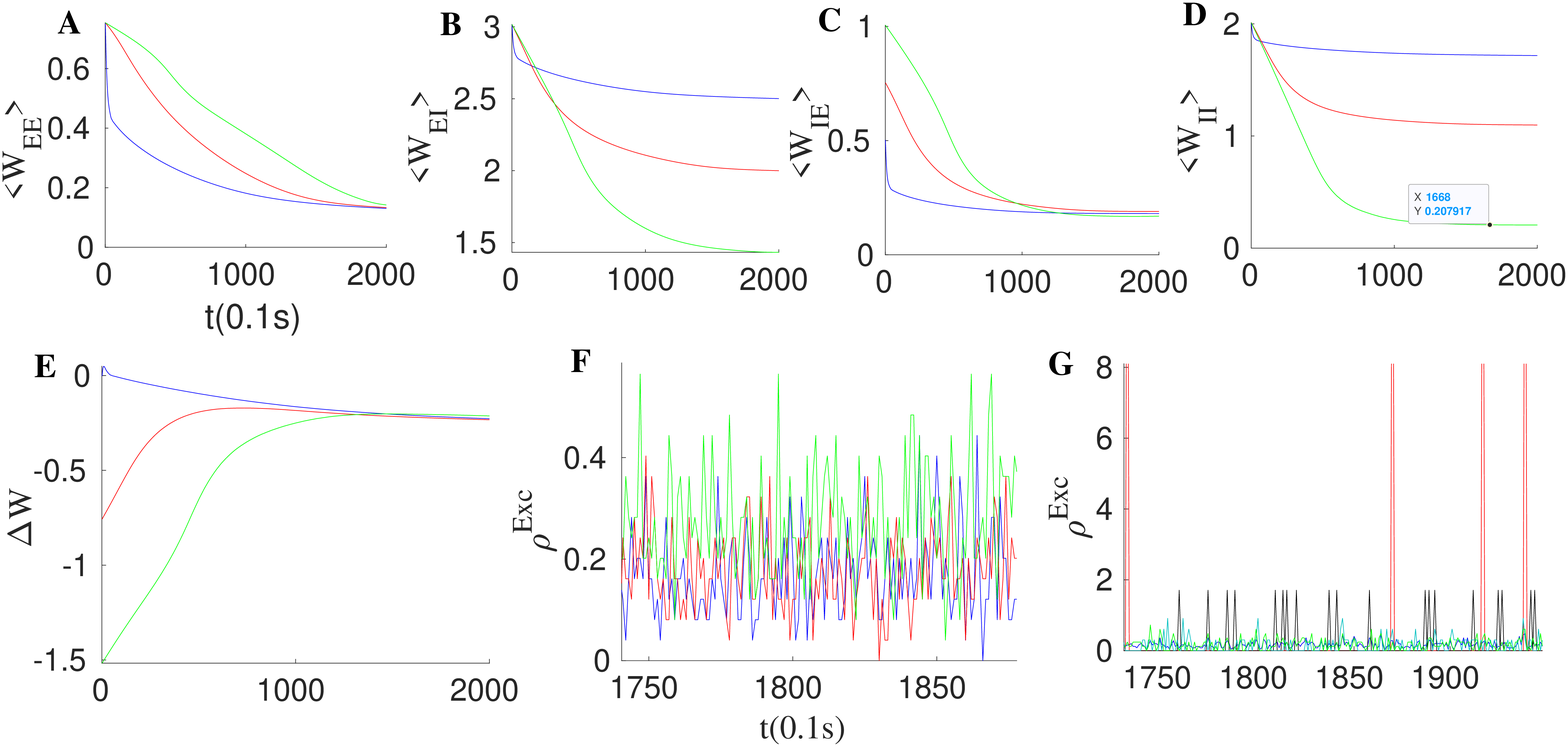}
		\caption[STDP tunes synaptic weights.]{Effect of  synaptic
                  plasticity on  network with three different initial weight
                  configurations when external excitatory input to both excitatory and inhibitory populations are of same magnitude  $\rho_
                  {Ext}$= 150Hz . (A-D) Evolution of  average synaptic
                  weights by STDP. (E) Change in the balance condition
                  by STDP. Slopes of the Exc. and the Inh. nullclines approach
                  each other under STDP in all three configurations. (F) The final state of the average neuron firing rates for these three networks lies below 1Hz. (G)
                  Network activity for different clusters of neurons with
                  different overall average inward synaptic weights. STDP
                  results in clusters with different overall connectivity
                  strengths and correspondingly different average
                  rates. }
                  \label{fig:1}
	\end{figure}	
Here, bars denote integrals of the kernels on the positive or negative real
lines. In the population of sparsely connected and sufficiently homogeneous
neurons, in terms of the number of connections of each neuron, and the regime
of asynchronous  homogeneous firing, i.e., when all the neurons fire with the
same average rate but with a random phase of firing between them, the average weights evolve as 
\begin{align}
& \dfrac {dw_{EE}}{dt} = \rho_E^2 \hat{K^E} + \rho_E w_{EE} (V_{Rexc} - \langle  V^E \rangle ) \overline{K_+^E \gamma^E} \nonumber \\
& \dfrac {dw_{EI}}{dt} =  \rho_E \rho_I \hat{K^I} - \rho_I w_{EI}(\langle  V^E \rangle - V_{Rinh} ) \overline{K_+^I \gamma^I} \nonumber \\
& \dfrac {dw_{IE}}{dt} =  \rho_E \rho_I \hat{K^E} + \rho_E w_{IE} (V_{Rexc} - \langle  V^I \rangle )\overline{K_+^E \gamma^E} \nonumber \\
& \dfrac {dw_{II}}{dt} =  \rho_I^2 \hat{K^I} - \rho_I w_{II} (\langle  V^I \rangle - V_{Rinh} ) \overline{K_+^I \gamma^I} \label{eq:12}
\end{align}

From the above equations, it is straightforward to see when $\hat{K^E} := \overline{K}_+^E - \overline{K} _-^E <0$ and $\bar{K^I} :=\overline{K}_+^I -\overline{K} _-^I  > 0 $, the stationary solutions satisfy :
\begin{equation} 
\dfrac {c_{EI}^{st}}{c_{EE}^{st}} =  \dfrac { \hat{K^I} \overline{K_+^E \gamma^E} }{ \hat{K^E} \overline{K_+^I \gamma^I}} = \dfrac {c_{II}^{st}}{c_{IE}^{st}}  \label{eq:13}
\end{equation}	
We take the proportion of inhibitory synapses to excitatory synapses to be equal for both excitatory and inhibitory neurons, i.e.  $\dfrac{k_{EI}}{k_{EE}} = \dfrac {k_{II}}{k_{IE}} $. The above condition brings the slopes of the excitatory and the inhibitory
nullclines close to each other (Equation (\ref{eq:6}))  leading to intersection in the semi-linear regime and proportionality of  excitation and  inhibition:
\begin{equation} 
\dfrac{\rho_I^{st}}{\rho_E^{st}} \approx \dfrac{c_{EE}}{c_{EI}} \label{eq:14}
\end{equation}	
As the system lies around the BT point, synaptic plasticity has a strong
  effect  when neurons are in a higher (here the linear) firing regime. At
this state, rates vary co-linearly  according to the above equation. On
the other hand, synaptic plasticity rules for $w_{II}$ and $w_{EI}$, i.e., the
second and fourth lines in equation (\ref{eq:13}), lead to a relation for stationary
weights in the form of $\dfrac{c_{II}}{c_{EI}} =
\dfrac{k_{II}\rho_I^{st}}{k_{EI}\rho_E^{st}}$. Comparing these last two equations, we arrive at
$k_{II} c_{EE}^{st} = k_{EI} c_{II}^{st}$. Assuming $k_{II} = k_{EI}$, the
mentioned relation adjusts the trace of the Jacobian at the fixed point in the
linear section to be near zero. Therefore, the plasticity rule and the dynamics of the near-linear regime stabilizes the system near the BT point.
Figure (\ref{fig:1}) shows that STDP brings the network of EI populations to the avalanche
regime. As STDP leads to an increase in the variance of the  weight distribution, some groups of  neurons become highly connected to each other while other groups show less overall connectivity strength. These groups of neurons will have different average rates as can be seen in Figure \ref{fig:1}G.

\subsection{Short-term plasticity tunes the network in a wide range of external input }
In the following, firstly, we will discuss the adaptive role of short-term
synaptic plasticity in bringing the network of the EI population to the
avalanche regime. Afterward, we will discuss how internal or external noise
close to the BT point can also cause the switch between the quiescent (Down)
and the low firing (Up) states. 
We will discuss that the Up-Down state transition by short-term depression can
be achieved either through a switch between bi-stable states or by bringing the system close to the BT point by dampening the overall excitation.
We use the model of \cite{Markram97} for short-term depression of the excitatory synapses reduces the
outgoing synaptic efficacy of excitatory synapses to an excitatory neuron in
case of a high rate of presynaptic activity. To model the STP effect, we
assume that the effective utility of the excitatory synapses of neuron
$j$ to the other neurons is proportional to the fraction of the available
synaptic resources $u$. Decrease of neurotransmitters at the synapses and
depression in release probability due to consecutive uses of neurotransmitters
in previous spikes of the presynaptic neuron are the sources of STP. We assume
by each spike of a presynaptic neuron, $u$  is reduced by the factor $qu$ and
then recovers with time constant $\tau_{STP}$ which is of order $100 ms$ to a few seconds. Therefore, synaptic efficacy of the postsynaptic synapse of neuron $j$ evolves as:
 \begin{equation} 
\dfrac{du_j}{dt} = \dfrac{1}{\tau_{STP}}(1-u_j) - qu_j \sum_k \delta(t-t_k^j)\label{eq:15}
\end{equation} 
Here, we just consider the short-term plasticity of synapses between
excitatory  neurons. This type of plasticity might occur in other types of
synapses as well, but we will not discuss this here. Because there exist
numerous input synapses and we have assumed homogeneous connectivity, each
neuron senses a large sample of the network activity and is connected with an
overall average weight with a small variance to the excitatory neuron
pool. Based on these assumptions and structural homogeneity,  we can write
down the dynamic of the average synaptic weights to the neuron $i$ in the
state of the network with excitatory population firing rate of
magnitude $\rho_E$ as :
\begin{equation}
\dfrac{dw_{EE}}{dt} = \dfrac{w_{EE}^0 -w_{EE} (t)}{\tau_{STP}} - w_{EE}(t)q\rho_E(t) \label{eq:16}
\end{equation} 
The rate equations for the EI population are of the  form:
\begin{align}
\dfrac{d\rho_{E}}{dt} &= - \dfrac{1}{\tau_m}(\rho_E(t) - f( \rho_{E}(t) , \rho_{I}(t) ,w_{EE}(t))) \nonumber \\
\dfrac{d\rho_{I}}{dt} &= -\dfrac{1}{\tau_m}(\rho_I(t) - g( \rho_{E}(t) ,  \rho_{I}(t) ) ) \label{eq:17}
\end{align} 	
Taking the time scale of short term plasticity to be much larger than the
EI-network activity decay time constant, i.e. $\tau_{STP} >> \tau_m$, we can
rewrite the dynamic in terms of fast, $d/dt^f$, and slow time, $d/dt^s$, evolution. Here, $t_f = t/\tau_m$ and $t_s = t/\tau_{STP}$. Defining $\mu = t_s/t_f$  and $\rho =$ 
$\begin{pmatrix}
  \rho ^E\\ 
  \rho ^I
\end{pmatrix}$, we arrive at
\begin{align} 
\dfrac{d\rho}{dt^f} &= - (\rho - f( \rho,w_{EE})) \nonumber \\
\dfrac{dw_{EE}}{dt^f} & = \mu ( w_{EE}^0 - w_{EE}) - q \tau _m w_{EE}\rho^E \label{eq:18}
\end{align}
This set of equations can have a stable fixed point or an oscillatory behavior. 
	\begin{figure}
			\centering
			\includegraphics[width=1\linewidth,trim={0cm, 0cm, 0cm, 0cm},clip]{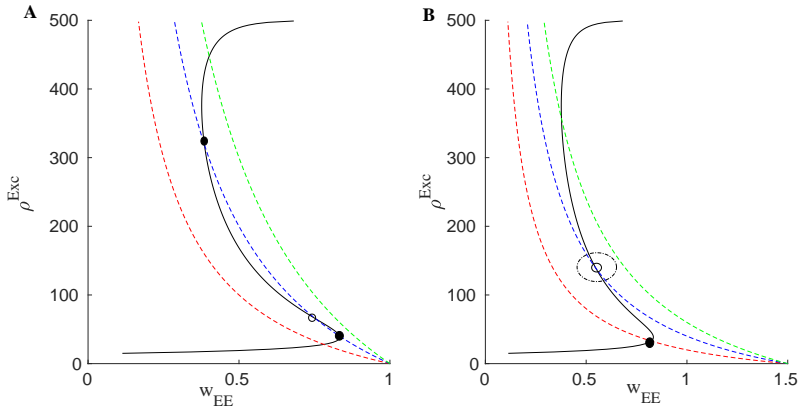}
		
		\caption[ Saddle-node and Hopf bifurcation in network with STP.]{Output excitatory rates as a function of $W_{EE}$ and the corresponding graphs for the average synaptic efficacy 
		$\langle W_{EE} \rangle _{St}$ at three values of $q$ (dashed
                red curve belongs to the largest and the dashed green curve is
                for the lowest value). Based on the value of $W_{EE}^0$, two
                different scenarios can occur. In (A) by decreasing
                $q$, through a saddle-node bifurcation stable and unstable
                fixed points appear at low and high values of the rates. In
                (B), with higher $W_{EE}^0$, by decreasing $q$
                after the Hopf bifurcation of the  low firing rate fixed point, an oscillatory solution for $(u , \rho_{out})$ emerges.} \label{fig:2}
	\end{figure}	
The average synaptic efficacy in the stationary state with the average excitatory rate $\rho_E^*$ is:
\begin{equation} 
\langle w_{EE} \rangle _{St} = \dfrac{w_{EE}^0}{1 +\tau q \rho_E^*} \label{eq:19}
\end{equation}
In case that there exists a fixed point or a stable limit cycle solution
around this point in the $(\rho_E ,\rho_I,\langle W_{EE} \rangle _{st})$ phase
space, the system might settle down at this solution (Fig.\ref{fig:2}A).  The
dynamics of the EI population near this region (with the slow-fast assumption) can be written as 
\begin{align}
\rho_{E}^{st} &= f( k_{EE}\rho_{E}^{st} , k_{EI} \rho_{I}^{st} ,\lambda_E^{Ex} , \lambda_I^{Ex}  ,\dfrac{w_{EE}^0}{1 +\tau q \langle \rho^{st} \rangle} ) \nonumber \\
\rho_{I}^{st} &= g( k_{EI}\rho_{E}^{st} , k_{II} \rho_{I}^{st} ,\lambda_E^{Ex} , \lambda_I^{Ex}  )  \label{eq:20}
\end{align}	
This mechanism is  effective to bring the system close to the BT point. Short
synaptic plasticity is a method of gain control that can bring the system from
a wide range of input and initial states to the low activity background
state. In Fig.\ref{fig:3}, we consider a system that is already tuned by STDP to the balanced state of weights (Eq. \ref{eq:13}) receiving various rates of external excitatory input to the excitatory population. In all cases, the system is initially away from the BT point. Without STP, the system shows a high firing rate oscillatory activity with an average rate of around $300Hz$. STP brings all of them closer to the avalanche regime. The average synaptic efficacy $\langle u \rangle $ in these cases does not oscillate significantly. As the system resides closer to the low firing regime through lowering the effective synaptic strength of Exc.-Exc. connections by STP, on a longer time scale STDP transforms the combination of weights to move the system to the avalanche regime by aligning the slopes of linearized nullclines. 

\begin{figure}
	\centering
			\includegraphics[width=1.1\linewidth,trim={0cm, 0cm, 0cm, 0cm},clip]{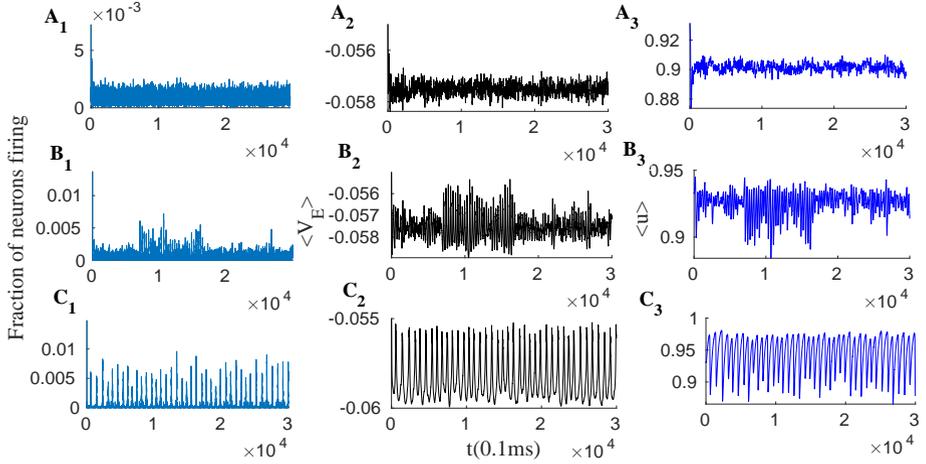}
		
		\caption[Population output in the network with STP]{EI
                  population with short term plasticity. $W_{EI} = 1.5, W_{II}=2, W_{IE}=0.75$, $W_{EE}^0=0.54$, $\rho_
                  {Ext}^{inh}$= 150Hz and $\rho_ {Ext}^{exc}$ = [380, 280, 
                  240]Hz. STP parameters are $q=0.3$, $\tau_{STP} =
                  10*\tau_{syn}$. The top panel is the system with the highest external input rate and the bottom panel is the one with the lowest. Left plots show the excitatory population rates, middle plots the population average membrane potential and right plots  the average excitatory synaptic efficacy $\langle u \rangle$.} \label{fig:3}
	\end{figure}

\begin{figure}[t]
			\centering
			\includegraphics[width=1\linewidth,trim={0cm, 0cm, 0cm, 0cm},clip]{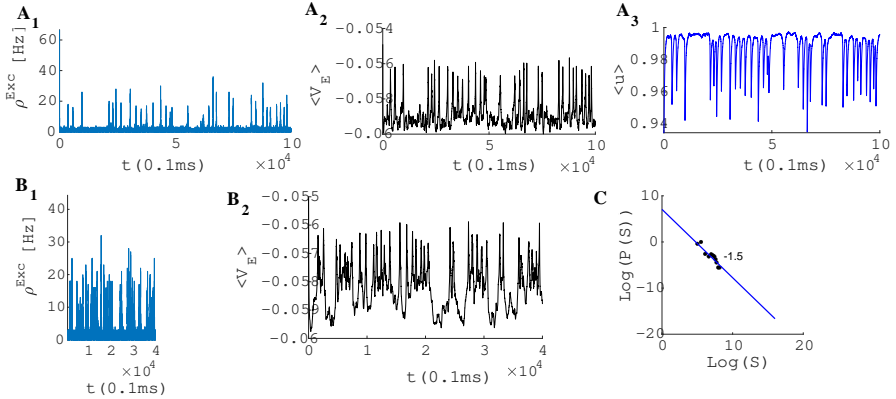}
		
		\caption[STP-UD]{(A)System with the same parameters values as in Fig.\ref{fig:3} except that here $\rho_{Ext}^{exc}=230 Hz$. (B) EI population with same parameters as in  Fig.\ref{fig:4}A but without STP. (C) Avalanche size
                  distribution in a log-log plot for the network in Fig.\ref{fig:4}A.} \label{fig:4}
	\end{figure}

 When the external input rate is tuned very close to the $BT$ point, where the quiescent fixed point and a low firing weakly (un)stable point lie close to each other, we see asynchronous avalanches of highly variable sizes  (see Fig. \ref{fig:4}A and Fig.\ref{fig:8}A). Without STP, we observe higher rates and less variable quiescent (Down) state time intervals (Fig.\ref{fig:4}B). Finite-size fluctuations kick the system out of the quiescent fixed point while STP plus fluctuations at the low rate fixed point drive the system back to the quiescent state. Average membrane potential as well as single neuron potentials in this case shows a transition between two levels (see Fig.\ref{fig:4}B$_2$).

\begin{figure}
			\centering
			\includegraphics[width=1\linewidth,trim={0cm, 0cm, 0cm, 0cm},clip]{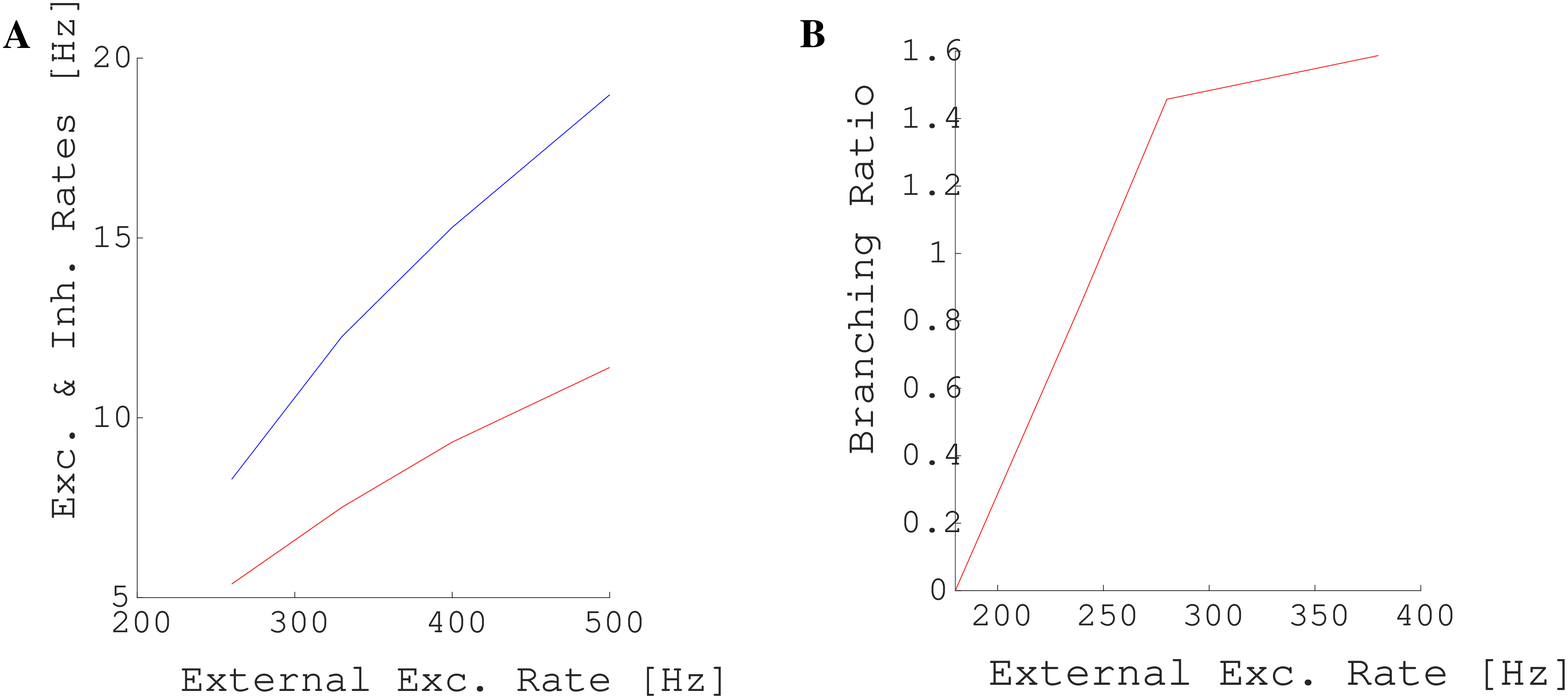}
		
		\caption[Avalanche characteristics and branching ratio in the
                final state of a network with STP.]{(A)
               The final
                  excitatory (red) and inhibitory (blue) output rates for the system in Fig.\ref{fig:3} and  Fig.\ref{fig:4}A . STP works as a gain control mechanism. (B) The branching ratio in the network states shown in  Fig.\ref{fig:3} and Fig.\ref{fig:4}A is defined as the average number of post-synaptic neurons of a single presynaptic neuron which set to fire by receiving the presynaptic input spike. Higher external rates set more neurons close to the threshold and thus the branching factor increases. When the inflection point of the steady membrane potential distribution passes the membrane threshold in the steady-state firing rate regime this increment rate slows down leading to the concavity of the branching factor curve.} \label{fig:5}
	\end{figure}

\begin{figure}
			\centering
			\includegraphics[width=1\linewidth,trim={0cm, 0cm, 0cm, 0cm},clip]{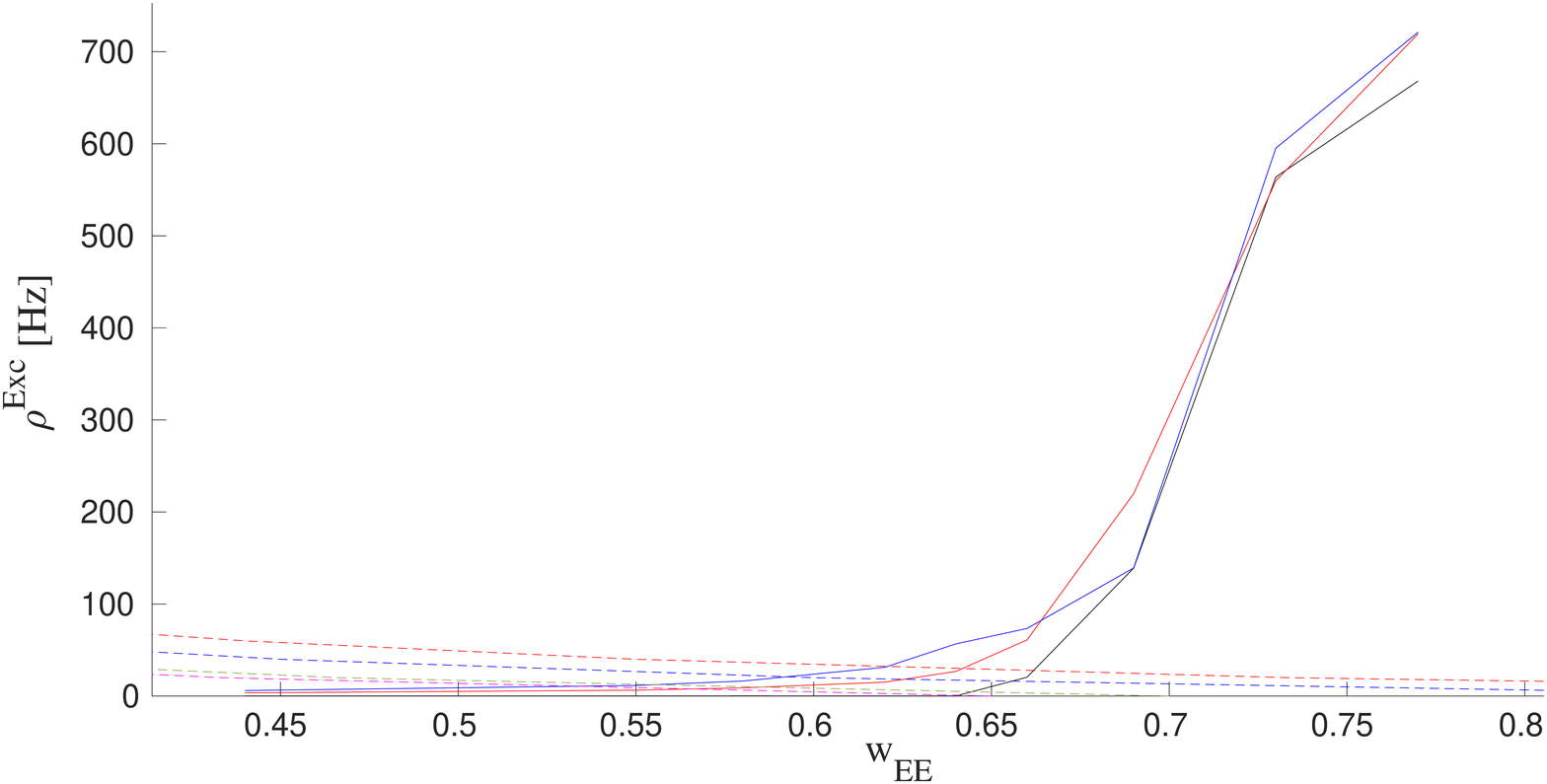}
		
		\caption[Input-output function vs. STP curves.]{Solid curves are the output rates for three different external input strengths vs. $W_{EE}$: Blue ($400Hz$), red ($310Hz$), and black ($220Hz$). Dashed curves show average stationary synaptic weight, $\langle W_{EE} \rangle $, in the network with STP with different maximum synaptic efficacies:  $W_{EE}^0 =$1.3 (red) , 0.9 (blue), 0.7 (green), and 0.65 (magenta). Intersections of the dashed and the solid curves are the fixed points of the EI network with STP for the corresponding control parameters. These fixed points are located in the low firing rate regime close to the avalanche region.} \label{fig:6}
	\end{figure}

Fig.\ref{fig:5}A shows excitatory and inhibitory stationary rates of the EI population subject to external rates in the range [200-500]Hz. As  can be seen, STP leads to low firing rate states and prevents overactivation.
Fig.\ref{fig:4}C shows avalanche size distribution in a log-log plot for
avalanches as in Fig.\ref{fig:4}A. The slope of the linear regression line
is very close to $-1.5$. The branching ratio for the final state of the system
is shown in Fig.\ref{fig:5}B. For $\rho_{Ext} = 230 Hz$, the branching ratio is
slightly less than one which is in agreement with our prediction in the
avalanche regime. 
	Moreover, Fig.\ref{fig:6} shows final output rates in the case of a stable fixed point for $u$  for three different values of the external excitatory rates and $W_{EE}^0$.
	
Another way that STP can cause a switch between two distinct firing states is
in the EI population which possesses bi-stability. In this case, change of $u$
can make each of the bi-stable nodes unstable while the system resides near
them. Decrease of $u$ in the up-state makes the up-state fixed point unstable
at some value of $u(t)$ (and accordingly $w_{EE}$). Therefore, the system will
jump to the remaining stable fixed point in a low or quiescent state. In the
very low firing regime (the quiescent state), $u$ will recover to its
asymptotic value, and the average synaptic weight increases towards
$w^0_{EE}$. If the quiescent state is unstable when $u$ approaches its maximum
value, we observe a transition to the high state.  Moreover, if the volume of
the basin of attraction of the quiescent fixed point is small, external and
internal noise can also induce the transition to the high rate fixed point and
the quiescent fixed point need not become unstable at $w^0_{EE}$.
For high values of $u$, the up-state fixed point is the stable point of the fast system but an unstable point of the slow one. Therefore, following the slow path up-state loses stability and the fast system remains with only a stable low fixed point. The trajectory of the slow $u$ is oscillatory in this case.
	\begin{figure}
			\centering
			\includegraphics[width=1\linewidth,trim={0cm, 0cm, 0cm, 0cm},clip]{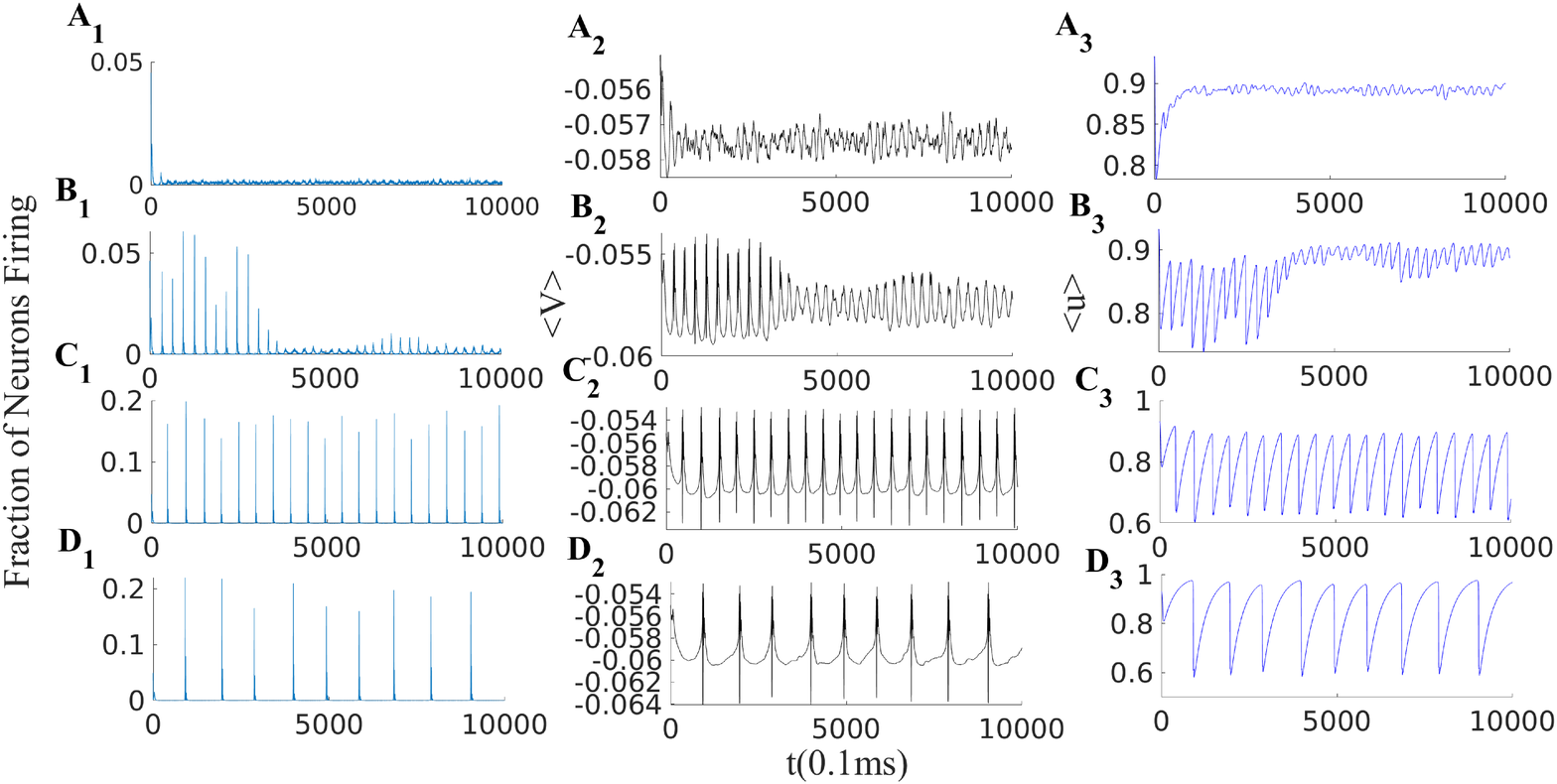}
		\caption[EI population with STP.(2)]{EI population with short
                  term plasticity. Network parmeters are : $W_{EI} = 2
                  ,W_{II}=2, W_{IE}=0.75$, $W_{EE}^0 =0.74$,
                  $\rho_{Ext}^{inh}$= 150 Hz, $\rho_{Ext}^{exc}$ =[ 400 ,330 ,
                  270, 225] Hz and STP parameters are: $q=0.4$ , $\tau_{STP} =
                  10*\tau_{syn}$ . Left plots: excitatory rates; middle plots:
                  average membrane potential; right plots: average synaptic efficacy $\langle u \rangle$. } \label{fig:7}
	\end{figure}	
Fig.\ref{fig:7} shows both ways that STP can produce synchronous avalanche behavior in
the system. When $W_{EE}=w^0$, the system is close to the constraints on the
alignment of the semi-linear segments of the EI-nullclines which results in
the presence of a high firing state as a unique fixed point of the system. In
the high input rate case, $\rho_{Ext} =400Hz$, corresponding to Fig.\ref{fig:7}A and the nullcline diagram of Fig.\ref{fig:8}C, due to STP, the system
moves from a high state of activity to a limit cycle solution at lower firing
rates. This final state is shown in Fig.\ref{fig:7}A and nullcline arrangements
in this state are depicted in Fig.\ref{fig:8}B. Here, there is an unstable source
in the linear branch sector which is surrounded by a limit cycle. Moreover,
oscillations in $\langle u \rangle$ have a low amplitude because of the
temporal averaging. On the other hand, the two bottom panels in Fig.\ref{fig:7} are
related to the situation of a switch between the high fixed point and the
quiescent node. As  shown in nullcline graphs in Fig.\ref{fig:8}C, at high synaptic
efficacy the high firing state is the only stable fixed
point, however, a high firing rate leads to a fast decline of the
synaptic efficacy which brings the system to the state with the nullcline map of Fig.\ref{fig:8}D which has a stable quiescent fixed point. The final state activity, in this case, is composed of avalanches with a high rate of firing in a short time window. Decreasing the $q$ factor can result in a longer up-state period. Also, $\langle u \rangle$ oscillates between two limits in these cases.

%		Nullcline plot for the system with the same parameters as in
%                  Fig.\ref{fig:7}A. (A) Corresponding to the system before
%                  STP with $ W_{EE}=0.74$. (B) The nullcline map after STP
%                  makes the high fixed point unstable and lowers the average effective Exc.-Exc. synaptic weights ($\langle W_{EE}\rangle$).
%
%
%Nullcline plot in the network with STP.(2)]{
%Nullcline plot for the system with same parameters as in Fig.\ref{fig:7}D. (A) Here, $W_{EE}=0.74$ which corresponds to the system being at the Up-state when the synaptic efficacy is fully recovered. (B) The nullcline map in the Down-state with low synaptic efficacy.

\begin{figure}
		
			\centering
			\includegraphics[width=1\linewidth,trim={0cm, 0cm, 0cm, 0cm},clip]{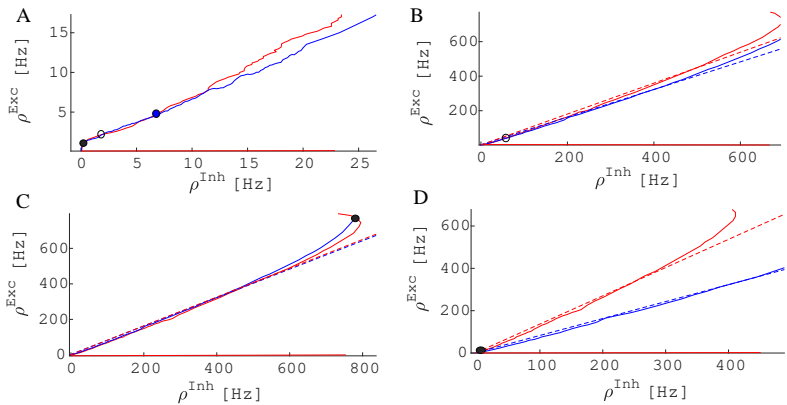}
		
		\caption[Nullclines plot in the network with STP.(1)]{ Excitatory (red) and inhibitory (blue) nullclines, i.e solutions to Eq.\ref{eq:4}) for excitatory and inhibitory rate, respectively. Dashed lines are the linearized nullcline approximation with slopes of $k\dfrac{c_{EI}}{c_{EE}}$ (Exc.) and  $k\dfrac{c_{II}}{c_{IE}}$ (Inh.) where k is a constant. STDP brings these slopes close to eachother (Eq. \ref{eq:13}.  (A) Nullclines at the Avalanche regime: The volume of the basin of attraction of the stable quiescent state is small and a weakly stable or unstable fixed point at the intersection in the semi-linear regime of nullclines is close to the origin. (B) System with only a saddle fixed point with a limit cycle solution with medium firing rates. (C) System with only a high firing fixed point. (D) System with only a stable quiescent state fixed point.} \label{fig:8}
	\end{figure}		
	
  In this particular case, necessary conditions for up to down transitions are:
\begin{align}
& k_{EE} W_{EE}^0 > k_{EI} W_{EI} \nonumber \\
& k_{EE} \dfrac{w_{EE}^0}{1 +\tau q \rho^{H}} < k_{EI} W_{EI}  \label{eq:21}
\end{align}
The first condition means that the slope of the excitatory nullcline is
smaller than the inhibitory one, which indicates a stable high firing fixed
point. The second condition states that at this high firing state the
stationary weight is not accessible before a stability loss. The slope of
nullclines increases by the decrease of effective $W_{EE}$ which causes the
high state to lose stability either through a Hopf or a saddle-node bifurcation.

 Finally, the plots in Fig.\ref{fig:7}B depicts the case where STP brings the system close to the BT point that shows low to medium size avalanches with higher variability. 
%	\begin{figure}
%		
%			\centering
%			\includegraphics[width=1\linewidth,trim={0cm, 0cm, 0cm, 0cm},clip]{./FrontiersPlots/UD_nulls2.eps}
%		
%		\caption[ } \label{fig:9}
%	\end{figure}		
In summary, transition from a quiescent state to a high firing state can be of
two distinct types. One way is that by increasing $W_{EE}$, the LF fixed point
and the unstable saddle move toward each other and in this way the basin of
attraction of the LF fixed point shrinks and noise can initiate the escape
from this fixed point to the high firing state. The other possibility is that
a fixed point losses stability through a Hopf bifurcation either before or after the emergence of a saddle-node in the middle branch.

\subsection{Avalanches and Turing instability in the interconnected network of EI populations}
The model of Wilson and Cowan for the dynamics of the spatio-temporal mean
fields of the excitatory and the inhibitory population rates, $E(x,t)$ and $I(x,t)$, in a 2D model of the cortex was introduced in the '70s. In this model at each point $x$, $E(x,t)$ is the density of the active neurons at time $t$ in a sphere of radius $\Gamma$ around $x$. 
Take $a$ as the average distance between a neuron and its neighbors and $\xi$
as the correlation length of the activity of neurons. Selecting $ a << \Gamma
< \xi $ guarantees that $E(x,t)$ is well defined and the number of neurons in
each block is large enough to take the limit w.r.t. the size of the local population. $E(x,t)$ is defined as  : 
\begin{align}
E(x,t) = \lim_{t \to 0} \lim_{N \to \infty}  \dfrac {n_{active} ( x  ,t+dt)}{N} \label{eq:22}
\end{align}
Where $n_{active} ( x  ,t)$ is the number of active neurons in the time window $[t,t+dt]$ and $N$ is the total number of neurons. Considering a high  density of neurons, we assume that fluctuations in the number of the active neurons around this mean value are negligible. Furthermore, to derive the dynamics of this field we have to write down the input-output relation for each population of neurons at position $x$. In the mean-field approximation, we assume that each neuron receives the same input, which implies that each neuron is connected to all other neurons, and the weights of connections depend only on the distance $x-x'$ between two neurons. This can be a valid approximation if the connections between adjacent neurons are dense and the heterogeneity in the network structure is minimal. 
 Now, assuming that neurons in each population relax to the non-active state in the absence of input with relaxation constant $\tau$, one can consider the following general equation for the activity field $E(x,t)$ :
\begin{equation} 
\tau \dfrac{dE(x,t)}{dt} = -E(x,t)+ ( k_e - r_e E(x,t)) * S_e( i_E(x,t))  \label{eq:23}
\end{equation}
$i_E(x,t)$ is the input to the excitatory population at $x$ at time $t$ which
 consists of both inhibitory and excitatory currents from both the
 self-activity of neurons in the population and other neurons in the adjacent
 population. $S_e$ is a nonlinear input-output function that in general
 depends on the response of the individual neurons, the  distribution of the membrane voltage, and the heterogeneity in weights.
In the following, like in Landau-Ginzburg modeling of a phase transition, we
will assume that  the fields $E(x,t)$ and $I(x,t)$ are very small so we can
perform an expansion as a  power series. Furthermore, since these are average
fields, we  assume that they vary slowly and smoothly in space, therefore, we
neglect fast fluctuations and write down a phenomenological model for the
current to the excitatory population as 
\begin{align}
i_E(x,t) = & \int_V [E(x',t)w_{EE}(|x'-x|)  +
  I(x',t)w_{EI}(|x'-x|)]  dV  +  i_E^{ext} \nonumber\\
= & E(x,t)\int w_{EE} dV +  \nabla E(x,t) \int (x'-x)w_{EE}(|x'-x|)dV \nonumber \\
  &+\dfrac{1}{2} \int (x'-x) H_E(x) (x'-x) w_{EE}(|x'-x|) dV  \nonumber \\
& +I(x,t)\int w_{EI}dV +  \nabla I(x,t) \int (x'-x)w_{EI}(|x'-x|)dV  \nonumber  \\ &+\dfrac{1}{2} \int (x'-x) H_I(x) (x'-x) w_{EI}(|x'-x|) dV  +  i_E^{ext} \nonumber \\
 =&2\pi E(x,t) \int r w_{EE}(r) dr + 2\pi I(x,t) \int r w_{EI}(r) dr \nonumber \\ &
 +\pi \nabla^2 E \int r^3 w_{EE}(r)dr +  \pi \nabla^2 I \int r^3 w_{EI}(r)dr + i_E^{ext} \nonumber \\&
  + \text{ higher order terms}  \label{eq:24}
\end{align}
We can expand $S^E$ around $ I_0^e = 2\pi E(x,t) \overline {rw_{EE}} + 2 \pi I(x,t) \overline{rw_{EI}}  + i_E^{ext}$ and write :
\begin{align} 
&\tau \dfrac{\partial E(x,t)}{\partial t} = -E(x,t) + ( k_e - r_e E(x,t)) [ S^E(I_0^e)  +  \pi S^{E'}_{\lvert I_0^E} ( \nabla ^2 E  \langle r^3 w_{EE} \rangle +  \nabla ^2 I  \langle r^3 w_{EI} \rangle) ]  \label{eq:25}
\end{align}
The same equation holds for the inhibitory field by expanding $S^I$ around  $ I_0^i =  2\pi E(x,t) \overline {rw_{IE}} + 2 \pi I(x,t) \overline{rw_{II}}  + i_I^{ext}$ : 
\begin{align} 
\tau \dfrac{\partial I(x,t)}{\partial t} = -I(x,t) +( k_i - r_i I(x,t)) [S^I(I_0^i)  +  \pi S^{I'}_{\lvert I_0^i} ( \nabla ^2 E \langle r^3 w_{IE}\rangle + 
  \nabla ^2 I  \langle r^3 w_{II}\rangle )] \label{eq:26}
\end{align}
Equations \ref{eq:25} and \ref{eq:26} are of the reaction-diffusion type for  the inhibitory and the excitatory fields in a two dimensional space. Defining $V(x,t) =$ 
$\begin{pmatrix}
  E(x,t)\\ 
  I(x.t)
\end{pmatrix}$, $D$ as the diffusion matrix and $f: R^2 \rightarrow R^2$ as
the gain function, one can rewrite the above equations in the following form:
\begin{equation}
\tau \dfrac{\partial V(x,t)}{\partial t} = D  \nabla ^2 V(x,t) + f(V(x,t)) \label{eq:27}
\end{equation}
The ODE part of this equation is the dynamic of a single EI population. The
corresponding low firing fixed point $(E_0, I_0)$ is stable in a specific
parameter regime. It usually loses stability either via a saddle-node or a
Hopf bifurcation which leads to either a region of bi-stability of low and
high firing states or the emergence of oscillations. However, still far away
from the bifurcation point since the diffusion matrix is not a scalar multiple
of the identity, Turing instabilities can occur in the system. It means that the homogeneous state of $[E(x,t),I(x,t)] =[E_0 , I_0]$ can become unstable. 
 In other words, when the diffusion coefficients of the inhibition and the excitation are sufficiently different, the homogenous steady state becomes unstable because of the diffusion.
To see this, noting that the above ODE has a homogeneous solution at $ V_0 =[E_0 , I_0] $, by linearization of the equation around the fixed point, i.e., $ V(x,t) = V_0 + v(x,t) $, and plugging in the ansatz $v(x,t) = e ^{ikx + \lambda (k)t } \psi $ we arrive at :
\begin{align} 
 \begin{pmatrix} -D_{EE} K^2 + \partial _{E\lvert V_0} f_E ,    - D_{EI}K^2 + \partial _{I\lvert V_0} f_E   \\ -D_{IE} K^2 + \partial _{E\lvert V_0} f_I , -D_{II}K^2 + \partial _{I\lvert V_0} f_I \end{pmatrix}  \psi=   \lambda \psi  \label{eq:28}
\end{align}
Assume $L$ to be the Jacobian of an isolated $EI$ population rate equation at $K=0$, then the condition for the stability of the fixed point reads as :
\begin{align} 
& \partial _{E} f_E  + \partial _{I} f_I < 0   \nonumber \\
 &det(L) = \partial _{E} f_E \partial _{I} f_I   -  \partial _{I} f_E \partial _{E} f_I  >  0   \label{eq:29}
\end{align}
 For the occurrence of a Turing instability at a critical $k_c$ the eigenvalues $\lambda(k) $ which satisfy the characteristic equation below should become positive for some real value of $k$:
\begin{align} 
\lambda ^2 + (( D_{EE}+D_{II}) k^2 - ( \partial _{E} f_E  + \partial _{I} f_I)\lambda +  R(k^2)  =0    \label{eq:30}     
\end{align}
with $R(k^2)$ defined as
\begin{align} 
R(k^2) = \alpha k^4 - k^2 ( D_{EE} \partial _{I} f_I   + D_{II} \partial _{E} f_E  - D_{IE} \partial _{I} f_E - D_{EI} \partial _{E} f_I  ) +  det(L)  \label{eq:31}  
\end{align}
in which $ \alpha := D_{EE}D_{II}-D_{EI}D_{IE}$. Since the coefficient of
$\lambda $ in the equation (\ref{eq:30}), i.e.,  the negative sum of the eigenvalues,
is positive, the necessary condition for the Turing instability is that
$R(k^2) < 0$  for some $k$ . When $\alpha >0 $, which is a reasonable
assumption as the inhibitory connections are more local than the excitatory
ones, if  $\beta := D_{EE} \partial _{I} f_I   + D_{II} \partial _{E} f_E  -
D_{IE} \partial _{I} f_E - D_{EI} \partial _{E} f_I  > 0$ then $R(k^2)$ can
become negative. In general $\dfrac{\beta}{\alpha}$ should be positive and
instability of the homogenous state occurs at the value $k_0^2 =
\dfrac{\beta}{2 \alpha}$ for which $R(k)$ becomes zero for the first time. In
the following, we consider the case of the local inhibitory connection and
therefore, we take $D_{II}$ and $D_{EI}$ to be very small.  Furthermore, we
assume that we are in the regime of positive $\alpha$, i.e., $D_{EE} > D_{IE}
$ and $ D_{II} \approx D_{EI} \ll D_{EE} $. With these assumptions, which are
intuitively valid based on the connectivity structure in the cortex, the condition for positive $\beta$ becomes:
\begin{align}
  D_{EE} \partial _{I} f_I  -  D_{IE} \partial _{I} f_E > 0  \label{eq:32}  
\end{align}
%\begin{figure}
%			\centering
%			\includegraphics[width=1\linewidth,trim={0cm, 0cm, 0cm, 0cm},clip]{./Article2Plots/STDpPop.eps}
%		
%		\caption[Final states of weights in the system with STDP satisfy condition for the Turing instability]{STDP changes values of connections' strengths in a way that sufficient conditions for emergence of Turing patterns, i.e. Eq.(18), are satisfied. (Left) $\langle W_{EE} \rangle$, (Middle) $\langle W_{II} \rangle$ and (Right)   $W_{EE} \mid W_{II} \mid - W_{IE} \mid W_{EI} \mid$, for three different initial sets of values. Network's paramteres evolve by STDP and at their final states satisfy conditions of Eq.(18).}
%	\end{figure}	
%  		
This holds when $D_{IE} $ and $\mid \partial _{I} f_E \mid$ are sufficiently
large which means that the excess inhibition will produce a larger effect on
the excitatory connection than the inhibitory one. This happens when the
average stationary state potential distribution for neurons is close to the
threshold, at which the effect of inhibition is larger than
excitation. Therefore, conditions for a Turing instability to occur are:
\begin{align} 
  & \mid \partial _{I} f_E \mid > \dfrac{D_{EE}}{D_{IE}}  \mid \partial _{I} f_I \mid  \nonumber \\
  &  \mid \partial _{I} f_I \mid  > \partial _{E} f_E \nonumber \\
  & \partial _ {E} f_I  > \dfrac{\partial _{E} f_E \mid \partial _{I} f_I  \mid }{ \mid \partial _{I} f_E \mid}  \label{eq:33}  
\end{align}
Since $D_{EE}$and $D_{IE}$ are proportional to $W_{EE}$ and $W_{IE}$  and the
partial derivatives in the first approximations are equal to the average
connection strength, the necessary conditions for the Turing instability reduce to 
\begin{align} 
& W_{IE} \mid W_{EI} \mid  > W_{EE} \mid W_{II} \mid \nonumber \\
& W_{EE} - \mid W_{II} \mid < 0   \label{eq:34}  
\end{align}
As can be seen in Fig.\ref{fig:1} and from the discussions following Eq(\ref{eq:13}), STDP leads
to the regime where the above conditions are satisfied.
\begin{figure}[t]
			\centering
			\includegraphics[width=1.1\linewidth,trim={0cm, 0cm, 0cm, 0cm},clip]{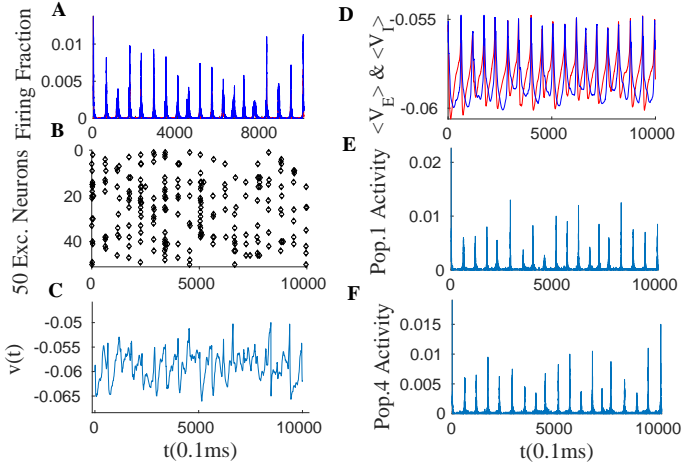}
		
		\caption[Simulation of 20 interconnected EI
                populations.]{Simulation result of 20 interconnected EI
                  populations each of size $N_E=10000$ arranged on a
                  ring. Average synaptic weights between two different EI
                  subnetworks decay with the phase difference of
                  them. Parameters of synaptic weights in each EI population
                  are   $W_{EE}=0.6$ , $W_{EI}= 2 ,W_{II}=2 , W_{IE}=0.75$
                  ,$\rho^E_{Ext}=230 Hz$ and $\rho^I_{Ext}=150 Hz$. (A)
                  The average firing rate of the whole network. (B)
                  Raster plot of the sub-population of 50  neurons in one
                  single EI subnetwork. Although neurons fire at avalanche
                  times, they don't fire in all of them, which leads to high
                  variability in their interspike interval
                  times. (C) The membrane potential of a single
                  neuron, which is in a constant transition between a state
                  close to the threshold and a state close to the resting
                  potential. (D) The average membrane potential of an
                  EI-population (red for Exc. and blue for Inh.). (E-F) The activity of two distinct EI subpopulations, showing high variability in sizes of avalanches in both of them.} \label{fig:10}
	\end{figure}	
Moreover, taking $ \partial _ {E} f_I \propto W_{IE}$  and $ \rho _{ext}$ as the free parameters, one can obtain the boundary of the Turing instability region defined by $ \min (R(k^2)) = 0$  at some value $k=k_c$. In this Turing instability region, stripe-like stationary patterns of activity with spatial frequencies close to $k = k_c$ appear in the system and the homogeneous solution becomes unstable.  
Besides Turing instabilities, it can also happen that the fixed point itself
loses its stability at $k=0$ through a Hopf bifurcation, where the real part of the eigenvalues of $L$ becomes zero:
\begin{align}
& \partial _{E} f_E  + \partial _{I} f_I = 0  \nonumber \\
 &det(L) = \partial _{E} f_E \partial _{I} f_I   -  \partial _{I} f_E \partial _{E} f_I  >  0   \label{eq:35}  
\end{align}
Furthermore, exactly at the BT point, we would also have $det(L) = 0$. Fig.\ref{fig:10} shows the activity of 20 interconnected EI-populations each operating close to the BT point. Overall activity in this system is of synchronous avalanches type. Up-down state transitions also become synchronized. 
We can model weakly interconnected EI populations in the avalanche regime which shows oscillation of frequency $\omega_{i}$ as pulse-coupled oscillators and therefore investigate conditions on synchronization and traveling wave solutions. This analysis is out of the scope of the current work.
Another approach consists in supplementing the macroscopic field equation with
an appropriate noise term to derive the mesoscopic equation. As can be seen
from Fig.\ref{fig:10}, the overall network activity is of avalanche type. This provides
again  evidence that avalanches are scale-free and occur in different temporal and spatial scales. However, our neural field model still lacks the internal finite-size fluctuation effects, inhomogeneities, and cross-correlation between individual neurons, and also inter-populations correlations. 
  	
  \subsection{Stochastic neural field }

\subsubsection{Finite size fluctuations in a single EI population}

So far we have analyzed mean-field models which were based on neglecting
finite system size, inhomogeneities in the synaptic connections, and
stochastic effects. Far from bifurcations of the mean-field (MF) equations, the
behavior of the average rates of the stochastic system follows predictions of
MF accurately. In this case, basins of attraction of the fixed points are
separated enough and stochastic effects do not lead to a change in the
macroscopic behavior of the system. However, close to the bifurcation points
of the macroscopic system, internal and external fluctuations can cause the
stochastic system to evolve differently from predictions of MF models. For example, it can cause transitions between different fixed points.	
	Let us consider a homogeneous network of size $N$ in which temporal and spatial
        variances in the firing rates of neurons are minimal. In this network,
        fluctuations in the finite system firing rates in the steady-state
        will be proportional to $ O(\dfrac{1}{\sqrt{N}})$. To model the
        finite-size stochastic effects, we need to write down dynamics of
        micro-state evolution that match the mean-field upon
        coarse-graining. As we have seen, the operating region of the EI
        population is around a low firing state where neurons fire with high
        variability of inter-spike intervals indicating that we can model
        their spiking as a Poisson process. In this regime of activity, the
        Poisson neuron assumption enables us to write down the microscopic
        evolution of a model neuron with two active and inactive states. The
        transition rate $\alpha$ between the active and the inactive state
        should  model vanishing of the postsynaptic potentiation, and the rate
        of inactive to active transition depends on the input and is therefore
        denoted
        by  $f(i)$. We want to model the system in the statistical homogenous state, in which the probability that a neuron fires depends only on the number of active neurons and therefore is the same for every neuron in the population. 
	
	In the sequel, we consider a population of $N_{E}$ excitatory and
        $N_{I}$ inhibitory neurons, in which neurons change their states
        independently. Let us take $P(E,I,t)$ as the probability density of
        the EI population being in a state with $E$ the number of active
        excitatory and $I$ the number of active inhibitory neurons at time $t$. The following master equation describes the microscopic evolution of the system:
 \begin{align} 
\dfrac{\partial P(E,I, t)}{\partial t} = & - \alpha [(EP(E,I,t)  +IP(E,I,t)]  \nonumber \\  &+  \alpha[ (E+1) P(E+1,I,t) + (I+1) P(E,I+1,t)] \nonumber \\
& +  (N_E - E + 1 )f( c_{EE} (E-1) ,  c_{EI}I) P(E-1,I,t) 
\nonumber \\ & - ( N_E -E)f( c_{EE} E, c_{EI}I) P ( E,I,t)  \nonumber \\
 &+(N_I - I + 1 )g(c_{IE}E ,  c_{II}(I-1)) P(E-1,I,t) 
 \nonumber \\ &- ( N_I -I)g( c_{IE}E, c_{II}I) P ( E,I,t)   \label{eq:36}  
 \end{align}  
in which $c_{xy} =\dfrac{K_{xy}}{N_x} w_{xy}$, with $K_{xy}$ being the number of incoming connections to a neuron in the population $x$ from the population $y$ , $f(.)$ and $g(.)$ are rates of inactive to active transtition for excitatory and inhibitory neurons,respectively.
We use the system size expansion method \cite{Vankampen07} for truncating the  moment hierarchy based on taking an ansatz on the order of the finite size fluctuation in the system.
Assuming fluctuations around the deterministic (average field) trajectory to be of order $O(N)$, we can rewrite our stochastic variables in terms of a deterministic and a fluctuating term as
\begin{equation}\label{eq:37} 
 E = N_E \rho _E + N_E^{1/2} \epsilon,  \qquad
 I = N_I \rho _I + N_I^{1/2} i 
\end{equation} 
where $\epsilon$ and  $i$ are representing fluctuations around the deterministic trajectories. Defining $P(E ,I ,t) =Q (\epsilon ,i ,t)$, we can rewrite the l.h.s. of the master equation in terms of the new parameters as 
 \begin{equation}\label{eq:38} 
 \dfrac{\partial P(E,I,t)}{\partial t} = \dfrac{\partial (\epsilon  ,i ,t)}{\partial t} - N_E^{1/2} \dfrac{d \rho_E(t)} {dt} \dfrac{\partial Q}{\partial \epsilon}   - N_I^{1/2} \dfrac{d \rho_I(t)} {dt} \dfrac{\partial Q}{\partial i}  
 \end{equation}
Defining ladder operators $Z_E f(E) = f(E+1)$ and $Z_E^{-1}f(E) = f(E-1)$  and expanding them in powers of $\epsilon $ , we arrive  :
\begin{align}
 Z_E = 1 + N_E^{-1/2}\dfrac{\partial}{\partial \epsilon} + \dfrac{1}{2} N_E^{-1}\dfrac{\partial^2}{\partial^2 \epsilon} + ... \nonumber\\
 Z_E^{-1} = 1 - N_E^{-1/2}\dfrac{\partial}{\partial \epsilon} + \dfrac{1}{2} N_E^{-1}\dfrac{\partial^2}{\partial^2 \epsilon} + ... \label{eq:39} 
 \end{align}
We define the same ladder operators for the inhibitory population states. Plugging all these equations into the master equation (\ref{eq:36}), we have:
 \begin{align} 
& \dfrac{\partial Q(\epsilon  ,i ,t)}{\partial t} - N_E^{1/2} \dfrac{d \rho_E(t)} {dt} \dfrac{\partial Q}{\partial \epsilon}   - N_I^{1/2} \dfrac{d \rho_I(t)} {dt} \dfrac{\partial Q}{\partial i} = \alpha ( Z_E - 1)[ (N_E \rho _E + N_E^{1/2}) \epsilon )Q] \nonumber \\
 &+ ( Z_E^{-1} - 1) [ N_E( 1-\rho _E - N_E^{-1/2} \epsilon )  *  f( c_{EE}N_E (\rho_E + N_E^{-1/2} \epsilon) ,  c_{EI}N_I(\rho_I + N_I^{-1/2} i) )  Q ]  \nonumber \\
  &+ \alpha ( Z_I - 1)[( N_I \rho _I + N_I^{1/2} i )Q(\epsilon  ,i ,t)]  \nonumber \\
 &+ ( Z_I^{-1} - 1) [ N_I ( 1-\rho _I - N_I^{-1/2} i ) * 
  g( c_{IE}N_E (\rho_E + N_E^{-1/2} \epsilon) ,  c_{II}N_I(\rho_I + N_I^{-1/2} i) )  Q ] \label{eq:40} 
 \end{align}
Expanding the inactive to active transition rates as 
\begin{align}
 f( c_{EE}N_E (\rho_E + N_E^{-1/2} \epsilon) , c_{EI}N_I(\rho_I + N_I^{-1/2} i) ) = &f( c_{EE}N_E \rho_E ,  c_{EI}N_I\rho_I ) \nonumber \\ &+   N_E^{-1/2} \dfrac{\partial f }{\partial \rho_E} \epsilon +N_I^{-1/2} \dfrac{\partial f }{\partial \rho_I} i + ...\label{eq:41} 
 \end{align}
Using the same expansion for ladder operators, we can sort the right and the left sides of equation (\ref{eq:40}) in powers of $N_E $ and $N_I$ . Equating terms of the order  $O(N_E^{1/2})$ and $O(N_I^{1/2})$ leads to the macroscopic equation:
\small
 \begin{align}
 - \dfrac{d \rho_E(t)} {dt} = \alpha \rho_E(t)  - (1-\rho_E(t))f( \kappa_{EE} w_{EE}N_E \rho_E  ,  \kappa_{EI}w_{EI}\rho_I )\nonumber \\
 - \dfrac{d \rho_I(t)} {dt} = \alpha \rho_I(t)  - (1-\rho_I(t))g( \kappa_{IE} w_{IE}N_I \rho_I  , \kappa_{II}w_{II}\rho_I )\label{eq:42} 
 \end{align}
 \normalsize
Equating terms of order $O(N^{0})$ leads to a linear FPE for $Q (\epsilon ,i ,t)$ of the  form:
 \begin{align}
 \dfrac{\partial Q(\epsilon  ,i ,t)}{\partial t} =& (\alpha - (1-\rho _E )\dfrac{\partial f }{\partial \rho_E} + f) \dfrac{\partial }{\partial \epsilon} \epsilon Q 
 + (\alpha - (1-\rho _I )\dfrac{\partial g }{\partial \rho_I} + f) \dfrac{\partial }{\partial i} i Q \nonumber \\
 &- (1-\rho _E )\dfrac{\partial f }{\partial \rho_I}  \dfrac{\partial }{\partial \epsilon} i Q 
 - (1-\rho _I )\dfrac{\partial g }{\partial \rho_E}  \dfrac{\partial }{\partial i} \epsilon Q \nonumber \\
 & + \dfrac{1}{2}(1-\rho _E )f \dfrac{\partial^2 }{\partial \epsilon ^2} Q
+\dfrac{1}{2}(1-\rho _I )g \dfrac{\partial^2 }{\partial i ^2} Q \label{eq:43}
 \end{align}
Defining matrices $A$ and $B$ as 
 \begin{align}
 \begin{pmatrix}
A_{11} & A_{12} \\
A_{21} & A_{22}
\end{pmatrix} &= \begin{pmatrix}
-\alpha + (1-\rho _E )\dfrac{\partial f }{\partial \rho_E} - f &  (1-\rho _E )\dfrac{\partial f }{\partial \rho_I}  \\
(1-\rho _I )\dfrac{\partial g }{\partial \rho_E}  & -\alpha + (1-\rho _I )\dfrac{\partial g }{\partial \rho_I} - g
\end{pmatrix} \nonumber \\
B &= \begin{pmatrix}
(1-\rho _E )f  & 0 \\
0 & (1-\rho _I )g
\end{pmatrix} 
 \end{align}
the amplitude of fluctuating term evolves as 
 \begin{align}
\dfrac{\partial}{\partial t} \begin{pmatrix}
\langle \epsilon \rangle \\
\langle i \rangle
\end{pmatrix}  =  \begin{pmatrix}
A_{11} & A_{12} \\
A_{21} & A_{22}
\end{pmatrix} \begin{pmatrix}
\langle \epsilon \rangle \\
\langle i \rangle
\end{pmatrix} 
 \end{align}
The covariance matrix $C = \begin{pmatrix}
Var(\epsilon)& Cov( \epsilon, i )\\
  Cov( \epsilon, i ) &  Var(i)
\end{pmatrix}$ satisfies :
\begin{align}
 \dfrac{\partial}{\partial t}C = AC + CA^t + B
\end{align}
If the determinant of $A$ is positive and its trace is negative at the
stationary point of the macroscopic equation (i.e., $A$ has two negative
eigenvalues), then the averages of the fluctuation terms go to zero. At the stationary point of the macroscopic equation, we  have:
\begin{align}
\begin{pmatrix}
A_{11} & 0 & A_{12} \\
0 & A_{22} & A_{21} \\
A_{21} & A_{12} & A_{11} +A_{22}
\end{pmatrix} \begin{pmatrix}
Var(\epsilon)_{st}\\
Var(i)_{st}\\
Cov(\epsilon , i)_{st} 
\end{pmatrix}
  =  -\dfrac{\alpha}{2} \begin{pmatrix}
\rho_E^{st}\\
\rho_I^{st}\\
0
\end{pmatrix}
 \end{align}
which has the  solution :
\begin{align} 
&Var(\epsilon)_{st} \approx c (( A_{11}A_{22}- A_{21}A_{12} + A_{22}^2) \rho_E +  A_{12}^2 \rho_I ) \nonumber \\
&Var(i)_{st} \approx c ((A_{11}A_{22}- A_{21}A_{12} + A_{11}^2) \rho_I + A_{21}^2 \rho_E) \nonumber \\
&Cov(\epsilon , i) _{st} \approx - c ( A_{11}A_{12} \rho_I+ A_{21}A_{22} \rho_E) \label{eq:48}
\end{align}
with $c = \dfrac{-\alpha}{2(A_{11}+A_{22})( A_{11}A_{22}- A_{21}A_{12})}$.

The average population rate and the fluctuation arround the macroscopic state are:
\begin{align} 
& \langle \dfrac{E}{N_E} \rangle = \rho_E \qquad  Var(\dfrac{E}{N_E}) =\dfrac{Var(\epsilon)}{N_E} \nonumber \\
& \langle \dfrac{I}{N_I} \rangle = \rho_I \qquad  Var(\dfrac{I}{N_I})=\dfrac{Var(i)}{N_I}  \label{eq:49}
\end{align}

\begin{figure}
			\centering
			\includegraphics[width=1\linewidth,trim={0cm, 0cm, 0cm, 0cm},clip]{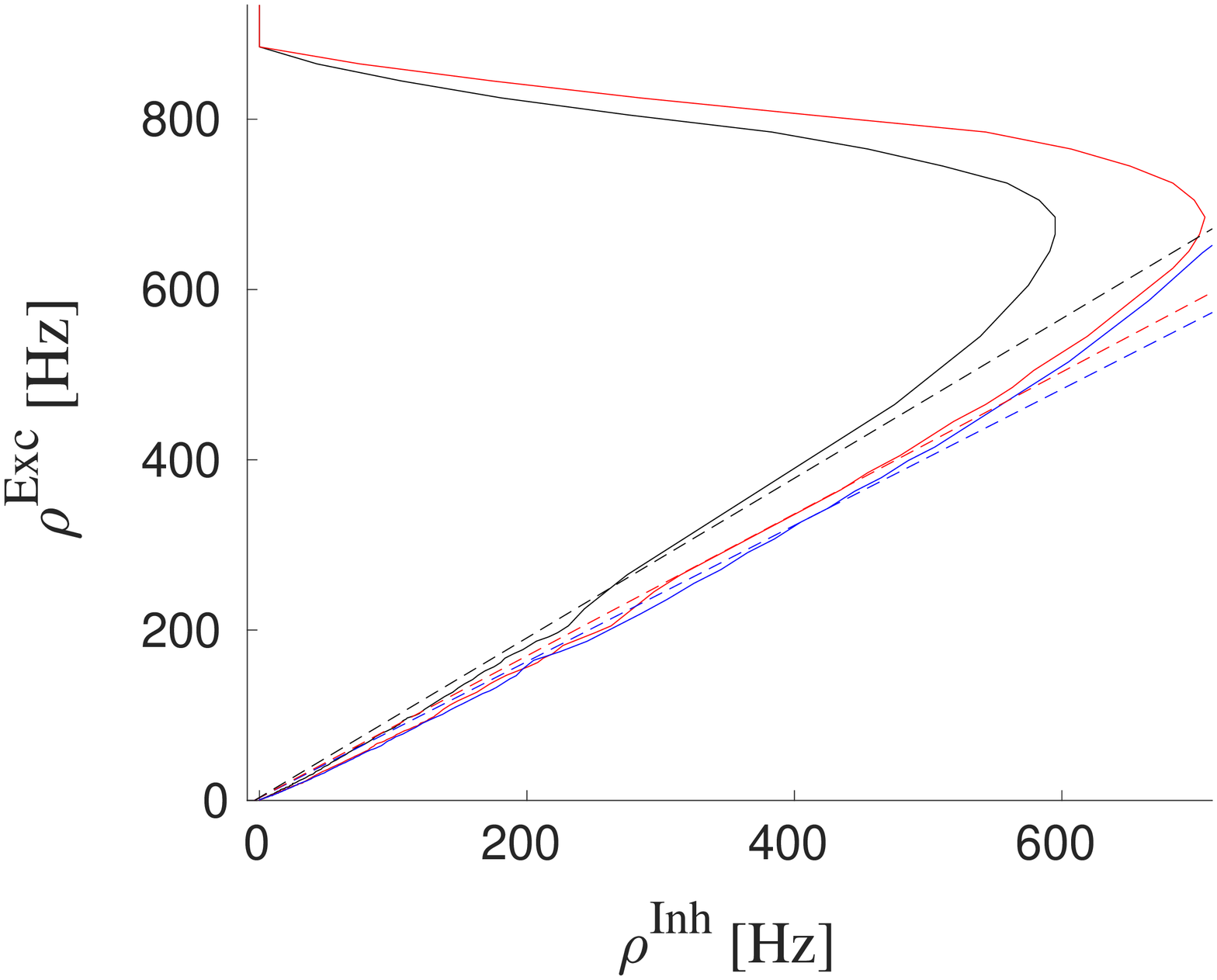}
		
		\caption[Nullcline graphs for the system in the Poisson firing
                regime.]{Nullcline graphs for the system with parameters:
                  $W_{EI} = 1.5 ,W_{II}=2 , W_{IE}=0.75,$ $\rho_{Ext}^I= 150Hz
                  , \rho_{Ext}^E = 280Hz $ ,$w_{EE}$ = 0.53(red) and
                  0.48(black). The blue curve is the inhibitory nullcline and dashed lines are linear approximations of the corresponding nullclines. Both excitatory nullclines have intersection with the inhibitory nullcline in the semi-linear section. } \label{fig:11}
	\end{figure}

\begin{figure}
			\centering
			\includegraphics[width=1\linewidth,trim={0cm, 0cm, 0cm, 0cm},clip]{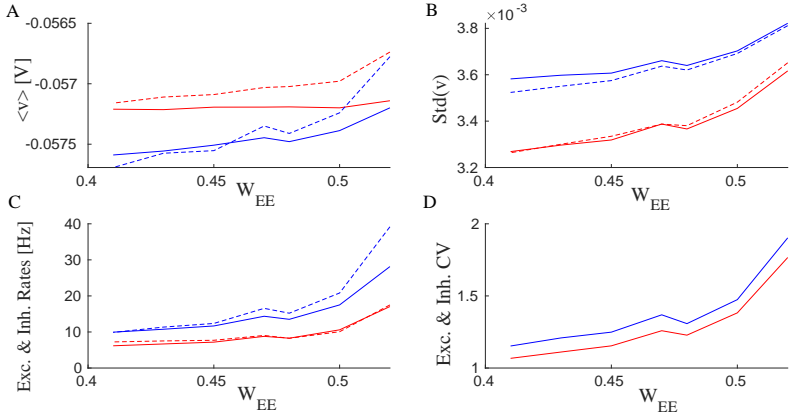}
		
		\caption[Characteristics of the network activity in the
                Poisson regime.]{Characteristics of the network activity for
                  systems with $W_{EI} = 1.5 ,W_{II}=2 , W_{IE}=0.75,$ $\rho_{Ext}^I= 180Hz
                  , \rho_{Ext}^E = 300Hz $ and 
                  $W_{EE} \in [0.41,52]$ which shows near Poisson firing and avalanche dynamics in smaller values of $W_{EE}$ in the mentioned value range. Red curves show excitatory
                  quantities and blue is for inhibitory ones. Dashed lines are
                  the prediction from the Poisson assumption and solid lines
                  are the simulation results. (A) Average membrane
                  potential. (B) The standard deviation of the
                  membrane potential. (C) Output rates. Here, dashed
                  lines are the firing rates derived from the Gaussian
                  approximation of the potential distribution based on values
                  of the average and the variance of the membrane potential in
                  the top panel of this figure. (D) CV of
                  interspike intervals in the simulation.}\label{fig:12}
	\end{figure}

\begin{figure}
			\centering
			\includegraphics[width=1\linewidth,trim={0cm, 0cm, 0cm, 0cm},clip]{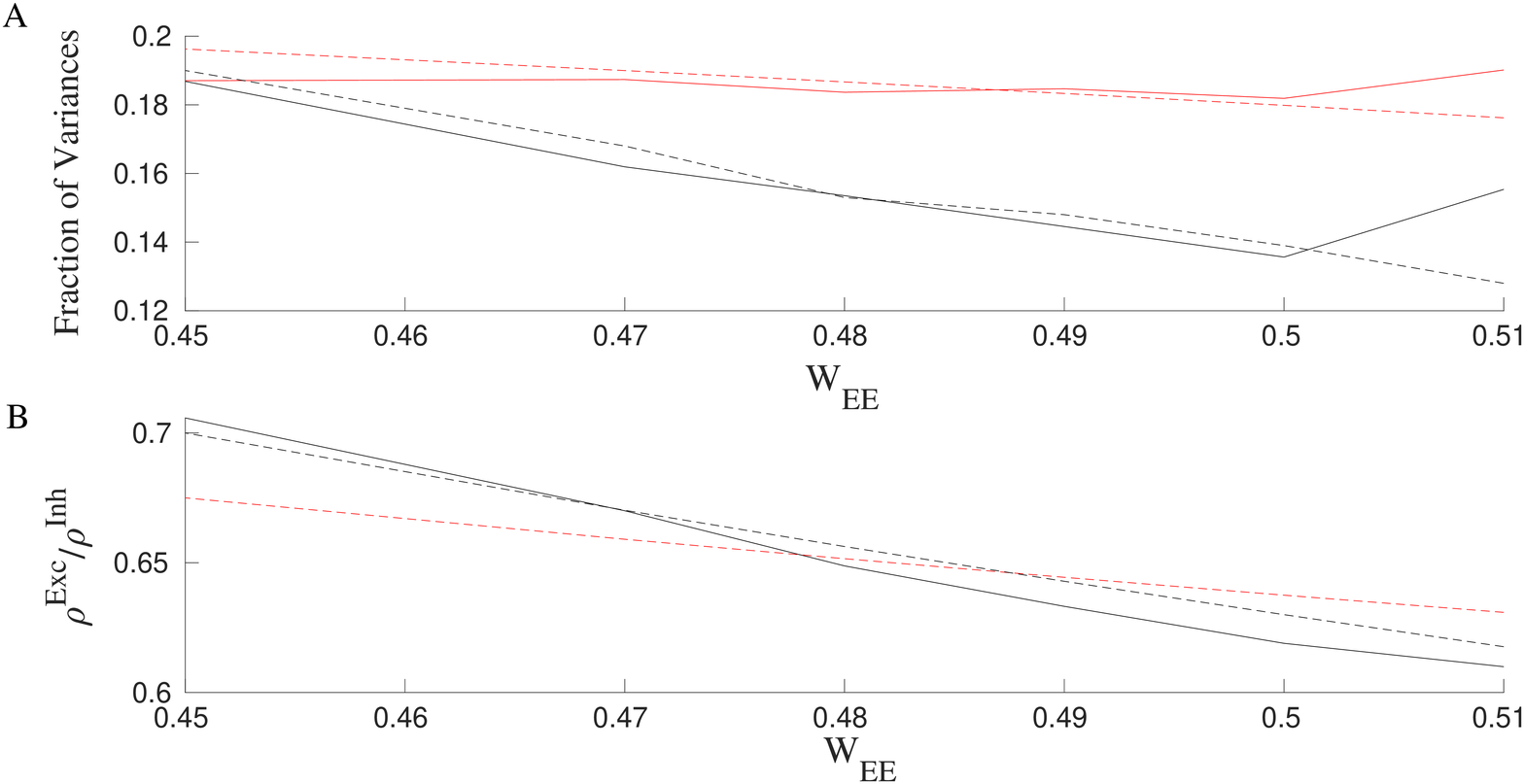}
		
		\caption[Variance of output population firing rate]{(A)
                  Ratio of var(E) to var(I) (red) and var(EI) to var(I)(black)
                  in the stationary state of the above mentioned
                  systems. Dashed lines are the approximation derived from the
                  Poisson assumption and solid lines are the simulation
                  results of the spiking neuron network. (B) The ratio of
                  the excitatory to the inhibitory stationary rates varies
                  vs. $W_{EE}$. The dashed line is $\dfrac{k_{EI}W_{EI}}{k_{EE}W_{EE}}$ and the solid line is the simulation result.} \label{fig:13}
	\end{figure}

From equations (\ref{eq:48}) and (\ref{eq:49}), it can be seen that close to the bifurcation of
the macroscopic equation, i.e., the BT point, where both trace and determinant
of the Jacobian are close to zero, fluctuation magnitudes  increase. Exactly
at the bifurcation point, the mentioned system size expansion fails because
the average of the noise term is unbounded, and therefore, we cannot assume
the fluctuating term in equation (\ref{eq:37}) to be of order $N^{1/2}$. Figures
(11-13) show characteristics of an EI population activity in the Poisson
regime. Fig.\ref{fig:11} is the nullcline graph related to two different strengths of
$W_{EE}$  both of them adjusting the system to have a stable fixed point on
the semi-linear part of the nullclines. The one with a higher $W_{EE}$ has a
higher firing rate fixed point. Fig.\ref{fig:12} shows how the average and the variance
of the membrane potential, the average rates, and the inter-spike interval CV
follow the prediction that neurons fire with Poisson statistics,
asynchronously and independently. The dashed lines in Fig.\ref{fig:12} show the
approximation with neurons  firing with Poisson statistics and independently
with the same rates that we observe in the simulation of the network. The
dashed lines in the bottom-left plot show the rate approximation with the
membrane potential distribution being Gaussian with the mean and the variance
as predicted in the top panel. Fig.\ref{fig:13}A shows that the variance and the covariance of the excitatory and the inhibitory rates in the simulation match the values derived from the microscopic model discussed above, i.e., equations (\ref{eq:48}) and (\ref{eq:49}).

\subsubsection{Stochastic neural field with a tuning mechanism to the critical state}
	
We have considered the situation where the inhibition effect in the network is local and we have seen that the system is tuned in a way that the inhibitory feedback in the local EI population balances the average excitatory current in a way that neuronal firing is fluctuation driven. 
If the effect of the inhibitory current is instantaneous, we can use
proportionality of excitatory and inhibitory currents to write down rate
dynamics of the excitatory population in terms of a stochastic field equation  when both inhibitory feedback and fluctuations are local. 
From equation (\ref{eq:14}), we know that near the BT point, there is a linear relation
between rates, i.e.,  $I \approx \dfrac{c_{EE}}{c_{EI}}
E$. Therefore,  the average current to the excitatory population close to the
BT point (Eq.\ref{eq:6}) can be written as $\gamma =c_{EE} E - c_{EI}I \approx 0
$. The second derivative in
the expansion of the gain function for the excitatory population from equation
(\ref{eq:27}) in the region of low activity is 
	\begin{align}
& \dfrac{1}{2}\dfrac{\partial^2 f}{\partial E^2} E^2 + \dfrac{\partial^2 f}{\partial EI} EI + \dfrac{1}{2}\dfrac{\partial^2 f}{\partial I^2} I^2
        \end{align}
        where we can use the following approximations for the gain function derivatives:	 
	\begin{align}
  \dfrac{\partial ^2 f}{\partial I^2} \propto  W_{EI}^2, \quad
  \dfrac{\partial ^2 f}{\partial E^2} \propto  W_{EE}^2, \quad
\dfrac{\partial f}{\partial I \partial E} \propto -W_{EE}W_{EI} 
\end{align}
By using the linear relation of the inhibitory and excitatory rates near the BT as the result of the projection of the dynamic to the slow manifold, we can replace inhibitory local field strength by a term linear in the local excitatory field.
Besides, fluctuations in the average population
activity, Eq.(\ref{eq:49}),  linearly depend on the rate. Therefore, we can write down the stochastic
field equation for the excitatory rate in the region of small $\gamma $:
	\begin{align} 
&\dfrac{\partial  E(x,t)}{\partial t} = \gamma E(x,t) + D \Delta E(x,t) - |u| E^2(x,t) + \psi (x,t)  \nonumber\\
	& \langle \psi (x,t) \psi (x',t') \rangle = 2  \sqrt {\dfrac{\sigma^2}{N}} E(x,t) \delta(x-x')\delta (t-t') 
	\end{align}
Here $u<0$ is the coefficient related to synaptic weights that can be
explicitly derived by assuming a certain form of the gain function and
proportionality of the rates. This stochastic partial differential equation
after appropriate rescaling $E(x,t) = \dfrac{\sigma}{|u|\tau}S(x,t)$ agrees
with the Langevin description of directed percolation which is of the following form with new transformed coefficients :
\begin{align} 
 \dfrac{\partial S(x,t)}{\partial t} = (\gamma '+D' \Delta) S(x,t)  - u' S^2(x,t) + \psi (x,t) \nonumber \\
 \langle \psi (x,t) \psi (x',t') \rangle = 2 u' S(x,t) \delta(x-x')\delta (t-t')
 \end{align}
  At $\gamma ' =0 $, the above system shows an absorbing state phase
  transition. Thus, from any active state the system relaxes by avalanches with a power-law size distribution to an inactive state.
  
\begin{figure}
 \centering
\includegraphics[width=0.5\linewidth,trim={0cm, 0cm, 0cm, 0cm},clip]{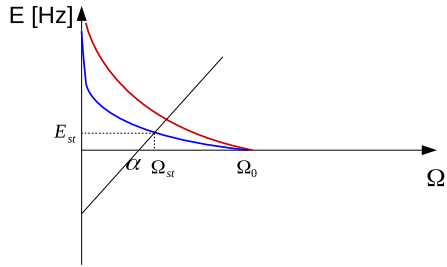}
\caption{Intersection of curves given by equations (\ref{eq:55}) and (\ref{eq:56}). Setting
  $\Omega_0$ to a value close to $\alpha$ by STDP
  and sufficient amount of synaptic depression leads to a stationary value of
  $\Omega_{st}$ very close to the critical value. The blue curve is associated
  to a higher value of $q$. } \label{fig:14}
			\end{figure}

   In an isolated EI population, external drive to the inhibitory and the
   excitatory population should be present to counterbalance the dissipation
   by the leaking currents and to thereby set the average membrane potential in these neurons at a state above the resting threshold. External
 excitatory input to the excitatory population is slightly higher than the
 external drive to the inhibitory population which leads to a slightly higher
 average membrane potential in the excitatory population. Furthermore, we can
 assume that the external spike train to each neuron is Poisson as well. The
 external drive by itself would not lead to significant firing in the individual neurons but the strengths of the internal connections between them are tuned in a way that bursts of activity occur in the excitatory population which is then followed by the inhibitory ones. The internal feedback inhibition is strong enough to kill the excitatory burst. In a slightly inhibition-dominated regime, we have sharp synchronous responses to the external input in a short time window. On the other hand, the network has a safe margin from an overly active state. In the absence of the input distinguished from random external noise, the system shows scale-free avalanches because of the maintenance of the inhibition-excitation balance.
 However, the external drive must compensate for the dissipation for the
 system tostay at or near the critical point. Without mechanisms like
 short-term plasticity, the external drive has be fine-tuned for the system to
 show criticality. However, short-term plasticity in a network in which
 synaptic weights are already near a slightly inhibition dominated regime
 broadens the range of the external drive strength which leads to critical
 avalanches. We can extend the short term synaptic depression of equation(8)
 to a continuum field equation by defining a field of excitatory synaptic
 efficacy  $\Omega(x,t) \propto \langle W_{EE} \rangle(x,t)$ with local
 dynamics of equation (\ref{eq:16}):
 \begin{align} 
&\tau_m \dfrac{\partial  E(x,t)}{\partial t} =   (-\alpha + \Omega(x,t)) E(x,t) + D \Delta E(x,t)  - |u| E^2(x,t) + \psi (x,t) \nonumber \\
&\dfrac{\partial \Omega(x,t)}{dt} =  \dfrac{1}{\tau_{STP}} (\Omega_0-\Omega) - q\Omega E(x,t) \nonumber \\
	&\langle \psi (x,t) \psi (x',t') \rangle = 2  \sqrt {\dfrac{\sigma^2}{N}} E(x,t) \delta(x-x')\delta (t-t') \label{eq:54}
	\end{align}
   Here $\alpha$ represents both the decay of  activity by leaky currents of
   the cells and the inhibition feedback which varies linearly with the
   excitatory rate. The dynamic excitatory synaptic strength brings the
   coefficient of the linear term to a value near zero (see Fig.\ref{fig:14}). This set of equations has a stationary synaptic efficacy solution of the  value
\begin{align} 
\Omega_{st} = \dfrac{\Omega_0}{1+q\tau_{STP}E_{st}}  \label{eq:55}
 \end{align}
On the other hand in the active phase, the stationary homogeneous rate is :
\begin{align} 
& E_{st} = \dfrac{-\alpha + \Omega_{st}}{|u|}  \label{eq:56}
 \end{align}
Assuming $|u|$ is a very small quantity and $E_{st}$ is also small in the low firing rate regime then $-\alpha + \Omega_{st} \approx 0$ and  $E_{st} = \dfrac{1}{q\tau_{STP}}(\dfrac{\Omega_0}{\alpha} -1) $.

Long-term synaptic plasticity tunes $\Omega_{0}$ so that the coefficient of
the linear term is close to zero and a moderate level of short-term depression
suffices to bring the system to the critical point. Altogether, equation
(\ref{eq:54}) is the description of an EI interconnected spiking neuron network tuned
to the critical point of balancing inhibition and excitation both by long-term
synaptic plasticity and short-term synaptic depression. The system wanders
around the phase transition point and shows avalanche dynamics with scale-free
size and time distribution, the proportionality of inhibition and excitation,
up and down state transitions of the membrane potential and population activity rates, and oscillations of order of $10Hz$ resembling ubiquitous alpha-band oscillations.

%%There are 5 heading levels
%
%\subsection{Level 2}
%\subsubsection{Level 3}
%\paragraph{Level 4}
%\subparagraph{Level 5}
%
%\subsection{Equations}
%Equations should be inserted in editable format from the equation editor.
%
%\begin{equation}
%\sum x+ y =Z\label{eq:01}
%\end{equation}

\section{Discussion}
In this work, we have proposed a self-organizing model for the cortical dynamics which tunes the system to the regime of low firing avalanche dynamics corresponding to the ongoing intrinsic activity in the cortex.
 We showed that long-term synaptic plasticity by STDP tunes the synaptic
 weights to achieve the internal balance of inhibition and excitation. This
 effect does not depend on the exact shape of STDP kernels. On the other hand,
 short-term synaptic depression can tune the system in response to the wide range of the strength of the external drive. 

In the vicinity of the bifurcation point, fluctuation effects are manifested
strongly and mean-field solutions do not adequately describe the dynamics. In
particular,  internal and external noises which were neglected in the
mean-field analysis can induce state transitions in the proximity of the saddle-node line. On the other hand, in the vicinity of the Hopf bifurcation line, fluctuations lead to higher variability in the amplitude of avalanches. In the vicinity of the BT point, by analyzing the linear stability of the quiescent fixed point, slow dynamics can be projected to the nonlinear stable manifold. 
We wrote down dynamics in terms of the excitatory field only  assuming a fast linear inhibitory feedback. The dynamical equation for the excitatory field matches the description of directed percolation. At the balance line, the coefficient of the linear term in the field equation vanishes which puts the system at the critical point of the percolation phase transition. Modeling the whole dynamic as a directed percolation of two dynamical variables is another possibility that we have not explored.

 In an open system, the external load has to be fine-tuned to compensate for
 the dissipation in order to remain at the critical point. Short-term
 depression of excitatory synapses allows this tuning for a wider range of
 external drives. We have not considered short-term plasticity for other types
 of synapses. However, we believe that under appropriate conditions on their respective strengths, we can observe the same qualitative regulation effects.

\section*{Conflict of Interest Statement}

The authors declare that the research was conducted in the absence of any commercial or financial relationships that could be construed as a potential conflict of interest.

\section*{Author Contributions}

M.E. and J.J. designed research. M.E. performed
research. M.E. wrote the manuscript. J.J. edited the
manuscript. All authors reviewed the manuscript, contributed to the article, and approved the submitted version.

\section*{Funding}
This study was funded by the International Max Planck Reseach School and the Max Planck Institute for Mathematics in Sciences in Leipzig.

\section*{Acknowledgments}
Open access funding provided by Max Planck Society. ME wants to thank International Max Planck Reseach School for funding his position during the research period.

\section*{Data Availability Statement}
The original contributions presented in the study are included in the article/supplementary material, further inquiries can be directed to the corresponding author.

\bibliography{Frontiersbib}

\end{document}